\documentclass[reprint,superscriptaddress,english,aps,prx,twocolumn,floatfix,showpacs,amsfonts,amssymb]{revtex4-1}
\usepackage{epsfig, graphicx,psfrag,amsmath,amssymb,float}
\usepackage[T1]{fontenc}
\usepackage[latin9]{inputenc}
\usepackage{amsmath}
\usepackage{amssymb}
\usepackage{bbold}
\usepackage{amscd}
\usepackage{bm}
\usepackage{psfrag}

\usepackage{babel}

\begin{document}
\title{Vanishing spin stiffness in the spin-$1/2$ Heisenberg chain for any nonzero temperature}

\author{J. M. P. Carmelo}
\affiliation{Department of Physics, University of Minho, Campus Gualtar, P-4710-057 Braga, Portugal}
\affiliation{Center of Physics of University of Minho and University of Porto, P-4169-007 Oporto, Portugal}
\affiliation{Beijing Computational Science Research Center, Beijing 100084, China}
\author{T. Prosen}
\affiliation{Department of Physics, FMF, University of Ljubliana, Jadranska 19, 1000 Ljubljana, Slovenia}
\author{D. K. Campbell}
\affiliation{Boston University, Department of Physics, Boston, MA 02215 USA}

\date{27 January 2015}

\begin{abstract}
Whether at zero spin density $m=0$ and finite temperatures $T>0$ the spin stiffness of the spin-$1/2$ $XXX$ 
chain is finite or vanishes remains an unsolved and controversial issue, as different 
approaches yield contradictory results. Here we provide an exact upper bound on the stiffness within a 
canonical ensemble at any fixed value of spin density $m$ and show that it is proportional to $m^2 L$
in the thermodynamic limit of chain length $L\to\infty$, for any finite, nonzero temperature. 
Moreover, we explicitly compute the stiffness at $m=0$ and confirm that it vanishes.
This allows us to exactly exclude the possibility of ballistic transport within the canonical ensemble for $T> 0$.
\end{abstract}


\maketitle

\section{Introduction}
\label{Introduction}

One-dimensional (1D) correlated lattice systems with $L$ sites show exotic spin transport properties 
at finite temperature $T>0$ whose nature has been a problem of long-standing both theoretical  
\cite{CZP-95,ZNP-97,Peres-99,Zotos-99,Gros-02,Kawa-03,Cabra-03,PSZL-04,ANI-05,Shastry-08,SPA-09,SPA-11,Tomaz-11,Znidaric-11,HPZ-11,Karrasch-12,Karrasch-13,Znidaric-13,Tomaz-13,Robin-14} and experimental \cite{Steiner-76,Motoyama-96,Thurber-01,Pratt-06,Branzoli-11,Yang-12,Yager-13} interest.
Often integrable quantum systems show dissipationless ballistic transport 
behavior. The real part of the corresponding conductivity as a function of the frequency 
$\omega$ and temperature $T$ has then a singular d.c. ($\omega=0$) contribution,
\begin{equation}
\sigma (\omega,T) = 2\pi\,D (T)\,\delta (\omega) +  \sigma^{reg} (\omega,T) \, .
\label{sigma-om}
\end{equation}

The stiffness, $D=D(T)$, is directly related to the time-average of the current-current correlation function as, 
\begin{equation}
D (T) = \frac{1}{2LT} \lim_{t\rightarrow\infty} \frac{1}{t}\int_0^t d t' \langle \hat{J} (t') \hat{J} (0)\rangle \, .
\label{D-T-gen}
\end{equation}
(Here and often in the following angle brackets denote thermal average.) Hence ballistic $D(T)> 0$ transport  means the correlation functions do not completely decay in time so their time-average is non-vanishing.

Integrable models are characterized by having a set of orthogonal commuting conserved quantities $\hat{Q}_j$ such that, 
\begin{equation}
\langle \hat{Q}_j \hat{Q}_{j'}\rangle = \delta_{j,j'}\langle \hat{Q}_j^2\rangle \, .
\label{QjQj}
\end{equation}
They provide an exact lower bound for $D$ encoded in an inequality due to Mazur \cite{Mazur}, 
\begin{equation}
D (T)\geq \frac{1}{2L}\sum_{j}{\langle\hat{J}\hat{Q}_{j}\rangle^2\over\langle \hat{Q}_j^2\rangle} \, .
\label{Mazur}
\end{equation}
Here the summation runs over all {\em linearly extensive} conserved quantities $\hat{Q}_j$ for which $\langle{\hat{Q}_j^2}\rangle\propto L$, local and
quasilocal \cite{Tomaz-11,Tomaz-13,Suzuki}.

The anisotropic spin-$1/2$ Heisenberg $XXZ$ chain with anisotropy parameter $\Delta\ge 0$ and exchange 
integral $J$  whose Hamiltonian reads,
\begin{equation}
\hat{H} = J\sum_{j=1}^{L}\left({\hat{S}}_j^x{\hat{S}}_{j+1}^x + {\hat{S}}_j^y{\hat{S}}_{j+1}^y + \Delta\,{\hat{S}}_j^z{\hat{S}}_{j+1}^z\right) \, ,
\label{HXXZ}
\end{equation}
is a paradigmatic example of an integrable strongly correlated system. 
Here $\hat{S}_j^{x,y,z}$ are components of the spin-$1/2$ operators at site $j=1,...,L$.
The related total spin operators, 
\begin{equation}
{\hat{S}}^{\tau} = \sum_{j=1}^{L}{\hat{S}}_j^{\tau} \, ;
\hspace{1.0cm}
{\hat{S}}^{\pm} = \sum_{j=1}^{L}{\hat{S}}_j^{\pm} \, .
\label{Sgenerators}
\end{equation}
where $\hat{S}_j^{\pm} = \hat{S}_j^x \pm i \hat{S}_j^y$ 
will play an important role in our study. 

Based on the lower bounds provided by Mazur's inequality, Eq. (\ref{Mazur}), it is known that the $XXZ$ 
chain exhibits ballistic spin transport at finite temperatures provided that the spin density $m$ is finite, $m\neq 0$. On the other hand, at zero spin 
density $m=0$ the spin current operator has  no overlap with any of the infinitely many local conserved quantities 
responsible for integrability, so that the use of Mazur's inequality is inconclusive.

Recently, exact high-temperature results by one of us
relying on the model's deformed symmetries and corresponding quasilocal 
conserved operators, i.e., nonlocal operators $Z$ for which $\langle Z^\dagger Z \rangle \propto L$ \cite{Tomaz-14,Pereira-14}, provided exact estimates for the spin stiffness at $m=0$ \cite{Tomaz-13}. 
Interestingly, for $\Delta = \cos (\pi/l)$, $l$ integer, this lower bound -- going to $0$ as $l\to\infty$ ($\Delta\to 1$) -- equals the spin-stiffness expression of Refs. \cite{Zotos-99,ANI-05}, which was derived by the original representation of the thermodynamic Bethe ansatz (TBA) \cite{Taka-AN}.

However, the isotropic point at $\Delta=1$ (the spin-$1/2$ $XXX$ model) \cite{Bethe} is the most experimentally relevant for
the spin-lattice relaxation rate and other physical quantities \cite{SPA-11,Motoyama-96,Thurber-01}.
It is also the case that poses the most challenging technical problems for theory. For instance,  
close to the isotropic point the numerical investigation of the spin stiffness expressions within the TBA \cite{Taka} calculating it from the eigenvalues of the Hamiltonian in a uniform vector
potential without the knowledge of matrix elements is difficult since the number of equations to solve
diverges \cite{Zotos-99}. Whether at $m=0$ and nonzero $T>0$ the spin stiffness vanishes or is finite remains an unsolved problem,
as different approaches yield contradictory results. 

On the one hand, several approaches (such as those used in the studies of
Refs. \cite{Karrasch-13,Karrasch-12,ANI-05,Cabra-03,Gros-02}) lead to a finite value for the spin stiffness. 
On the other hand, the studies of Ref. \cite{SPA-11} show that transport
at finite temperatures is dominated by a diffusive contribution, the spin stiffness being very small or zero.
Such studies exclude the large spin stiffness found in Ref. \cite{ANI-05} by a phenomenological
method that relies on a spinon and anti-spinon particle basis for the TBA. 
The infinite-temperature result of Ref. \cite{Znidaric-11} (based on a nonequilibrium open system approach) suggests that the
$XXX$ chain exhibits anomalous sub-ballistic spin transport. The TBA 
results of Refs. \cite{Zotos-99,Peres-99} find a vanishing spin stiffness for zero spin density.
The more recent results of Ref. \cite{HPZ-11} reached the same conclusion by combining several techniques. 

In this paper we provide new insights that partially resolve the above unsolved problem concerning the spin stiffness for spin $1/2$ $XXX$ chain in the 
thermodynamic limit (TL) $L\to\infty$. Specifically, we show that it vanishes exactly as $m\to 0$ within the canonical ensemble 
for fixed total spin projection $S^z$ (note that $m=-2S^z/L$), including $S^z=0$, at least as fast as, 
\begin{equation}
D_{S^z}(T) \le \frac{4c}{T} \frac{(S^z)^2}{L} = \frac{c}{T} m^2 L \, , 
\label{Dcm0}
\end{equation}
where $c$ is an $L,S^z,T$--independent constant. A similar result is also reached for a canonical ensemble near 
the fully polarized sector of maximal spin density $m=1$,
\begin{equation}
D_{S^z}(T) \le \frac{c'}{T} (1-m)^2 L \, , 
\label{Dclinem1}
\end{equation}
where $c'$ is another constant. 

That our results {\it partially} resolve the stiffness behavior of the spin-$1/2$ $XXX$ model as $m \rightarrow 0$
stems from their leaving out, marginally, the grand canonical ensemble
in which $\langle m^2\rangle = {\cal O}(1/L)$. However, our study relies onto an exact stiffness upper
bound whose derivation involves a large overestimation of the elementary currents carried by the
energy and momentum eigenstates. Hence accounting for the usual expectation of the equivalence of 
the canonical and grand canonical ensembles in the TL, we expect that our results remain valid in the 
latter grand canonical case.

The remainder of the paper is organized as follows.
The $S^z=0$ spin stiffness and the introduction of some operator algebra 
useful for the studies of this paper are the issues addressed
in Sec. \ref{D-algebra}. In Sec. \ref{ME-QN} the spin current operator matrix elements 
that contribute to the stiffness are expressed in terms of the quantum numbers that label the
energy eigenstates. The optimization of the spin current value in each 
reduced subspace spanned by energy eigenstates with fixed $S$ and remaining quantum-number values
is the problem studied in Sec. \ref{optimizing}. In Sec. \ref{two} two exact spin stiffness upper bounds
that follow from the optimization of the spin current in each reduced subspace are derived.
Finally, the concluding remarks are presented in Sec. \ref{concluding}.

\section{The $S^z=0$ spin stiffness and some useful operator algebra}
\label{D-algebra}
        
We consider the spin-$1/2$ $XXX$ chain Hamiltonian with periodic boundary conditions,
which is that given in Eq. (\ref{HXXZ}) at the isotropic point, $\Delta =1$,
\begin{equation}
\hat{H} = J\sum_{j=1}^{L} \hat{\vec{S}}_j \cdot \hat{\vec{S}}_{j+1} \, .
\label{Hchain}
\end{equation}
The key to our analysis will be to exploit the $SU(2)$ symmetry, $[\hat{H},\hat{S}^\tau]=0$, $\tau\in\{x,y,z\}$, with
the spin operators $\hat{S}^\tau$ and $\hat{H}$ given in Eqs. (\ref{Sgenerators}) and (\ref{Hchain}), respectively.
The energy eigenstate's spin and spin projection are denoted by $S$ and
$S^{z}=-(N_{\uparrow} -N_{\downarrow})/2$, respectively. 
For the so-called highest/lowest-weight-states (HWSs/LWSs) of the $SU(2)$ algebra we have $S=S^z$/$S=-S^z$.

The $z$-component of the {\it spin current operator} can be written as,
\begin{equation}
\hat{J} = -i\,J\sum_{j=1}^{L}(\hat{S}_j^+\hat{S}_{j+1}^- - \hat{S}_{j+1}^+\hat{S}_j^-)  \, .
\label{c-s-currents}
\end{equation}

The LWSs and the non-LWSs generated from them used in our analysis are energy and momentum eigenstates.
They are as well eigenstates of $(\hat{\vec{S}})^2$ and $\hat{S}^z$ with eigenvalues $S(S+1)$ and $S^z$,
respectively. We thus denote all 
$2^{L}$ energy and momentum eigenstates by $\vert l_{\rm r},S,S^z\rangle$. Here $l_{\rm r}$ stands for all quantum 
numbers other than $S$ and $S^z$ needed to specify an 
energy and momentum eigenstate, $\vert l_{\rm r},S,S^z\rangle$. 
The non-LWSs are generated from the corresponding
$n_s= S + S^z=0$ LWS $\vert l_{\rm r},S,-S\rangle$ as,
\begin{equation}
\vert l_{\rm r},S,S^z\rangle = 
\frac{1}{\sqrt{{\cal{C}}}}({\hat{S}}^{+})^{n_s}\vert l_{\rm r},S,-S\rangle \, ,
\label{nonLWS}
\end{equation}
where 
\begin{equation}
{\cal{C}} = (n_s!)\prod_{j=1}^{n_s}(\,2S+1-j\,) \, ;
\hspace{0.5cm} n_s= 1,...,2S \, .
\label{CnonLWS}
\end{equation}

Within the canonical ensemble description of a 1D correlated system, the spin stiffness for $T>0$ can be written in terms of
a summation over current matrix elements between energy eigenstates as \cite{Shastry-08},
\begin{equation}
D (T) = {1\over 2TL}\sum_{\nu} p_{\nu}
\sum_{\nu' (\epsilon_{\nu}=\epsilon_{\nu'})} 
\vert\langle\nu\vert\hat{J}\vert\nu'\rangle\vert^2 \, , \hspace{0.25cm} T \geq 0 \, .
\label{DT-gen}
\end{equation}
Here the Boltzmann weight and the partition function read $p_{\nu}=e^{-\epsilon_{\nu}/T}/Z$ 
and $Z = \sum_{\nu}  e^{-\epsilon_{\nu}/T}$, respectively. 

For large $L$ there are two temperature regimes: (i) $T$ smaller and (ii) $T$ larger than the energy eigenstate 
level spacing \cite{ZNP-97}. In the limit $L\rightarrow\infty$, the temperature 
regime (i) shrinks to $T=0$, while the temperature region (ii) includes all of $T>0$. 

In regime (i), ($T=0$), $D (0)$ is finite and is given by $D(0)=J/(2\pi)$ \cite{Shastry-90}.
On the other hand,  in regime (ii) ($T>0$), the stiffness expression, Eq. (\ref{DT-gen}), simplifies 
in the TL, {\it  provided} that one chooses the energy eigenstates to be also momentum eigenstates. 
Accounting for the vanishing in the TL of the persistent currents \cite{Fowler-92}, 
one finds \cite{ZNP-97,Ginsparg-90} from the exact cancellation of some contributions by
summing over momentum $k$ and $-k$ subspaces, the result that the expression of $D (T)$
in terms of energy and momentum eigenstates $\vert l_{\rm r},S,S^z\rangle$ involves only current 
expectation values. The general expression, Eq. (\ref{DT-gen}), then simplifies to, 
\begin{equation}
D (T) = {1\over  2T L}\sum_{\nu} p_{\nu}\vert\langle\nu\vert\hat{J}\vert\nu\rangle\vert^2 
\, , \hspace{0.25cm} T > 0 \, .
\label{DT-gen-Tlarger0}
\end{equation}
Within the canonical ensemble at fixed value of $S^z$ we can therefore exactly 
define the spin Drude weight $D_{S^z}(T)$ at finite temperature and in the TL 
in terms of our representation, Eq. (\ref{nonLWS}), for the energy and momentum
eigenstates as,
\begin{equation}
D_{S^z} (T)  =  {1\over 2 T L}\sum_{l_{\rm r} } \sum_{S=\vert S^z\vert}^{L/2} p_{l_{\rm r},S,S^z} 
\vert\langle l_{\rm r},S,S^z\vert\hat{J} \vert l_{\rm r},S,S^z\rangle\vert^2 \, .
\label{D-T}
\end{equation}
Here the Boltzmann weights $p_{l_{\rm r},S,S^z}$ and the partition function $Z_{S^z}$ should be defined with respect to sums over all 
${L\choose L/2-S^z}$ energy and momentum eigenstates with fixed $S^z$. In this and 
all following expressions for the stiffness the sums over $S$ always increase in  {\em steps} of $1$, whereas $S^z$ and
$S$ have to be integers (half-odd integers) for even (odd) $L$.

The commutators, 
\begin{eqnarray}
\left[\hat{J},\hat{S}^{\pm}\right] & = &
\left[\hat{S}^{z},\hat{J}^{\pm}\right] = \pm \hat{J}^{\pm} 
\, ; \hspace{0.25cm}
\left[\hat{J}^{\pm},\hat{S}^{\mp}\right] = \pm 2\hat{J}
\nonumber \\
\left[\hat{J},\hat{S}^{z}\right] & = & 0
\, ; \hspace{0.25cm}
\left[\hat{J},(\hat{\vec{S}})^2\right]  = \hat{J}^{+}\hat{S}^{-} - \hat{S}^{+}\hat{J}^{-} \, ,
\label{comm-currents}
\end{eqnarray}
which follow directly from $SU(2)$ algebra for the spin operators, play a major role in our study. Here in addition to its $z$-component, Eq. (\ref{c-s-currents}), the other two $SU(2)$ 
symmetry operator components $\hat{J}^{\pm}$ of the current  read,
\begin{equation}
\hat{J}^{+} = (\hat{J}^{-})^{\dag} = 2i\,J\sum_{j=1}^{L}(\hat{S}_j^+\hat{S}_{j+1}^z - \hat{S}_{j+1}^+\hat{S}_j^z) \, .
\label{Jpm}
\end{equation}

The $S>0$ LWSs $\vert l_{\rm r} ,S,-S\rangle$ and the $S=S^{z}=0$ states (which are simultaneously 
LWSs and HWSs $\vert l_{\rm r},0,0\rangle$) used
in our operator algebra manipulations obey the well-known transformation laws
$\hat{S}^{-}\vert l_{\rm r} ,S,0\rangle = 0$ and
$\hat{S}^{+}\vert l_{\rm r},0,0\rangle = \hat{S}^{-}\vert l_{\rm r},0,0\rangle = 0$,
which follow straight-forwardly from the corresponding $SU(2)$ symmetry operator algebra. 

To derive useful exact relations involving the current expectation values $\langle l_{\rm r} ,S,S^z\vert\hat{J}\vert l_{\rm r} ,S,S^z\rangle$ that
appear in the $T>0$ spin stiffness expression, Eq. (\ref{D-T}), we consider in the following 
some more general current matrix elements between energy, momentum, and $(\hat{\vec{S}})^2$ eigenstates,
$\langle l_{\rm r} ,S,S^z\vert\hat{J}\vert l_{\rm r}',S',S^{z}\rangle$, given by,
\begin{eqnarray}
& & \langle l_{\rm r} ,S,S^z\vert\hat{J}\vert l_{\rm r}',S',S^{z}\rangle
= {1\over\sqrt{{\cal{C}}{\cal{C}}'}}
\nonumber \\
& \times & \langle l_{\rm r} ,S,-S\vert ({\hat{S}}^-)^{n_s}\hat{J} 
({\hat{S}}^+)^{n_s'}\vert l_{\rm r}',S',-S'\rangle \, .
\label{matrixele}
\end{eqnarray}
Here the normalization constants are given in Eq. (\ref{CnonLWS}),
and we have accounted for the vanishing of the commutator $[\hat{J},\hat{S}^{z}] = 0$,
Eq. (\ref{comm-currents}), so that the current operator connects only states with the same $S^z$ 
value. For $l_{\rm r}=l_{\rm r}'$ and $S=S'$, $\langle l_{\rm r} ,S,S^z\vert\hat{J}\vert l_{\rm r}',S',S^{z}\rangle$
refers to the current expectation values in Eq. (\ref{D-T}). 

We start by considering a class of current 
matrix elements $\langle l_{\rm r} ,S,S^z\vert\hat{J}\vert l_{\rm r}',S,S^{z}\rangle$
between states with the same arbitrary $S\geq 1/2$ and $S^{z}$ values. The
following general result is valid for $S \geq 1/2$,
\begin{equation}
\langle  l_{\rm r} ,S,S^z\vert\hat{J}\vert  l_{\rm r}',S,S^z\rangle = -{S^z\over S}
\langle  l_{\rm r} ,S,-S\vert\hat{J}\vert  l_{\rm r}',S,-S\rangle \, , 
\label{currents-gen}
\end{equation}
where $S^z = -S + n_s$ and $n_s = 1,...,2S$. It is obtained by
combining the systematic use of the commutators given in Eq. (\ref{comm-currents}) with the
above state transformation laws.
The calculations to reach Eq. (\ref{currents-gen}) are relatively easy for non-LWSs whose generation 
from LWSs involves small $n_s=S-S^z$ values. The calculations become lengthy as the $n_s$ 
value increases, but they remain straightforward. 

We shall now and in the ensuing sections study separately the two cases, $S^z=0$ and $S^z\neq 0$. 
Indeed, it is important not to restrict ourselves only to the case of strictly $S^z=0$ (requiring $L$ to be even), 
which may be sensitive to certain pathologies 
and thus to consider the spin stiffness for any finite fixed value of $S^z$ in the TL.

Analysis of the matrix elements, Eq. (\ref{currents-gen}), reveals that the 
$l_{\rm r} = l_{\rm r}'$ and $S^z = 0$ current expectation values $\langle  l_{\rm r} ,S,0\vert\hat{J}\vert  l_{\rm r},S,0\rangle$ all vanish for 
$S\geq 1/2$. However, we also need such current expectation values for
$S=S^z=0$. Those are the particular case, $l_{\rm r} = l_{\rm r}'$, of the
general matrix elements $\langle l_{\rm r} ,0,0\vert\hat{J}\vert l_{\rm r}',0,0\rangle$, 
which in the following are shown to vanish. Such  matrix elements connect the energy eigenstates 
$\vert  l_{\rm r} ,0,0\rangle$ and $\vert  l_{\rm r}',0,0\rangle$ which are both LWSs and HWSs. 
It follows from Eq. (\ref{comm-currents}) that the current operator $\hat{J}$, Eq. (\ref{c-s-currents}), may be written 
as the commutator $\hat{J} = {1\over 2}[\hat{J}^{+},\hat{S}^{-}]$.
Thus the current matrix elements $\langle l_{\rm r} ,0,0\vert\hat{J}\vert l_{\rm r}',0,0\rangle$ can be written as,
\begin{eqnarray}
& & \langle l_{\rm r} ,0,0\vert\hat{J}\vert l_{\rm r}',0,0\rangle =
\nonumber \\
&  & {1\over 2}(\langle l_{\rm r} ,0,0\vert\hat{J}^{+}\hat{S}^{-}\vert l_{\rm r}',0,0\rangle
- \langle l_{\rm r} ,0,0\vert\hat{S}^{-}\hat{J}^{+}\vert l_{\rm r}',0,0\rangle) .
\label{matrixele00}
\end{eqnarray}

That this expression vanishes is readily confirmed by applying the above-stated
transformation laws. A similar result holds for all 
matrix elements of the form $\langle  l_{\rm r} ,S,0\vert\hat{J}\vert  l_{\rm r}',S+\delta S,0\rangle$
where $S \geq 0$ and $S' = S + \delta S\geq 0$, which are found to vanish unless $\delta S = \pm 1$.
Hence all $S^z = 0$ current expectation values $\langle  l_{\rm r} ,S,0\vert\hat{J}\vert  l_{\rm r},S,0\rangle$ vanish for 
$S\geq 0$, so that Eq. (\ref{D-T}) yields $D_{S^z=0}(T) = 0$ for $T > 0$.

\section{Matrix elements in terms of the quantum numbers that label the energy eigenstates}
\label{ME-QN}

Since vanishing spin density $m=0$ may in the TL also be approached by any finite fixed $S^z$, or fixed window of $S^z$ values,
and then letting $L\to\infty$, we must carefully estimate $D_{S^z}(T)$ for $S^z\neq 0$.
Expressing current expectation values in terms of expectation values in LWSs, using
the matrix-element relations of Eq. (\ref{currents-gen}) for $l_{\rm r} = l_{\rm r}'$ and $S\geq 1/2$, 
we obtain,
\begin{eqnarray}
D_{S^z} (T) & = & {(2S^z)^2\over 2 L T}\sum_{S=\vert S^z\vert}^{L/2}\sum_{l_{\rm r}} 
p_{l_{\rm r},S,S^z}
\nonumber \\
&\times& {\vert\langle l_{\rm r},S,-S\vert\hat{J} \vert l_{\rm r},S,-S\rangle\vert^2\over (2S)^2} \, .
\label{D-all-T-simp}
\end{eqnarray}

Out of the $\sum_{2S=0\,({\rm integers})}^{L}\,{\cal{N}}(S) = 2^{L}$ energy eigenstates, there are ${\cal{N}}(S) = (2S+1)\,{\cal{N}}_{singlet} (S)$ states
for a given $S$. For $S>0$, each such state is populated by a set of $2S$ spins $1/2$ that
participate in the $2S+1$ multiplet configurations and a second set of $L-2S$ 
spins $1/2$ that are within the
\begin{equation}
{\cal{N}}_{singlet} (S) = {L\choose L/2-S}-{L\choose L/2-S-1} 
\label{NsingletS}
\end{equation}
spin-singlet configurations, which involve $(L-2S)/2$ spin-singlet pairs. Straightforward arguments then imply that only the first
group of $2S$ spins contributes to the spin current $\langle\hat{J}\rangle = {1\over 2}\langle[\hat{J}^{+},\hat{S}^{-}]\rangle$ of the above states.

Within the TBA solution of the model, part of the degrees of freedom of the $(L-2S)/2$ spin-singlet pairs
are distributed over a set $\{M_n\}$ of configurations each with
$M_n$ $n$-pair configurations. Here $n=1,2,...$ is the number
of spin-singlet pairs. Consistently, 
\begin{equation}
{1\over 2}(L-2S) = \sum_{n=1}^{\infty}n\,M_n \, .
\label{Nsingletpairs}
\end{equation}
For $n>1$ the spin-singlet pairs of a $n$-pair configuration are bound within it. 

The model Hamiltonian, Eq. (\ref{Hchain}), is solvable by the Bethe ansatz, the corresponding general
Bethe-ansatz equation being of the form \cite{Bethe,Taka},
\begin{equation}
2\arctan (\Lambda_j) = q_j + {1\over L}\sum_{\alpha\neq j}2\arctan\left({\Lambda_j-\Lambda_{\alpha}\over 2}\right) 
\hspace{0.20cm} {\rm mod}\,2\pi \, .
\label{gen-Lambda-BA}
\end{equation}
Here the $\alpha =1,...,(L-2S)/2$ summation is over the set of occupied $q_{\alpha}$ quantum numbers
$q_j = {2\pi\over L}I_j$ and the occupancies of the related quantum numbers $I_j$ (defined modulo $L$) 
such that $j = 1,...,(L+2S)/2$ label the energy eigenstates. They are half-odd integers for $(L+2S)/2$ 
even and integers for $(L+2S)/2$ odd. 

The LWS Bethe-ansatz wave functions formally vanish when two rapidities $\Lambda_j$ and $\Lambda_{j'}$ become
equal. This property suggests that simply choosing $\alpha =1,...,(L-2S)/2$ distinct quantum numbers
$q_{\alpha}$ among the set of $j = 1,...,(L+2S)/2$ allowed quantum numbers $q_j$, which gives a dimension,
\begin{equation}
{(L+2S)/2\choose (L-2S)/2} \, ,
\label{dim-LS}
\end{equation}
would allow the reconstruction of all $2^L$ energy eigenstates that span the model Hilbert space. 

However, this expectation is misleading. Indeed, only some of the solutions to the general
Bethe-ansatz equation, Eq. (\ref{gen-Lambda-BA}), are in terms of real rapidities $\Lambda_j$.
There also exist solutions involving groups of complex rapidities \cite{Bethe,Taka}. In the present TL they
have the general form \cite{Taka},
\begin{equation}
\Lambda_j^{n,l} = \Lambda_j^n + i (n+1-2l) \, , \hspace{0.5cm} l = 1,...,n \, ,
\label{Lambda-jnl-ideal}
\end{equation}
where $j = 1,...,M_n^b$ and the number $M_n^b\geq M_n$ is defined below.
Use of these solutions in the general Bethe-ansatz equation, Eq. (\ref{gen-Lambda-BA}),
leads to a set of $n = 1,...,(L-2S)/2$ coupled integral equations. These are the TBA equations 
given in the following. As confirmed below, the new set of quantum numbers associated
with such equations allows the reconstruction of the set of $2^L$ energy eigenstates that span the full Hilbert space. 

For $n=1$ the rapidity, Eq. (\ref{Lambda-jnl-ideal}), is real and refers to a single unbound spin-singlet pair. The imaginary part that emerges 
for $n>1$ is associated with the binding of the corresponding $n$ spin-singlet pairs. Often in the
literature the complex rapidity, Eq. (\ref{Lambda-jnl-ideal}) for $n>1$, is called a $n$-string. Moreover, 
the number of bound pairs $n$ is often called the string length and the real part of the rapidity, $\Lambda_j^n$, the string center. 

Importantly, the $n$-pair configurations prevail under the finite-system complex rapidity deviations 
\cite{Vladimirov-84,Essler-92,Isler-93,Takahashi-03,Caux-07} from
their ideal form, Eq. (\ref{Lambda-jnl-ideal}). As discussed in Appendix \ref{string}, such deviations do not
change the spin currents carried by energy eigenstates without $n>1$ bound pairs. Indeed, 
they only affect the finite-system currents carried by energy eigenstates with $n>1$ bound pairs. 
This is consistent with the finite-system deviations from the complex-rapidity ideal strings, Eq. (\ref{Lambda-jnl-ideal}), 
not contributing to the thermodynamics provided that $T>0$ and $m\neq 0$ 
\cite{Caux-07,Wiegmann-83}.

After some algebra, one finds that the use of rapidities of the form, Eq. (\ref{Lambda-jnl-ideal}),
in the general Bethe-ansatz equation, Eq. (\ref{gen-Lambda-BA}), leads to the $n = 1,...,(L-2S)/2$ TBA equations
\cite{Taka}, which within the momentum-distribution functional notation used in this paper read,
\begin{equation}
k_n (q_j) = q_j + {1\over L}\sum_{(n',j')\neq (n,j)}M_{n'} (q_{j'})\,\Theta_{n\,n'} (\Lambda_j^n-\Lambda_{j'}^{n'}) \, .
\label{gen-Lambda}
\end{equation}
Their solutions of define the rapidities real part, $\Lambda_j^n$. In them,
\begin{equation}
k_n (q_j) = 2\arctan (\Lambda_j^n/n) \, ,
\label{kn-gen-Lambda}
\end{equation}
and $\Theta_{n\,n'}(x)$ is an odd function of $x$ given by,
\begin{eqnarray}
& & \Theta_{n\,n'}(x) = \delta_{n,n'}\Bigl\{2\arctan\Bigl({x\over 2n}\Bigl) 
\nonumber \\
& + & \sum_{l=1}^{n -1}4\arctan\Bigl({x\over 2l}\Bigl)\Bigr\} 
\nonumber \\
& + & (1-\delta_{n,n'})\Bigl\{ 2\arctan\Bigl({x\over \vert\,n-n'\vert}\Bigl)
\nonumber \\
& + &  2\arctan\Bigl({x\over n+n'}\Bigl) 
\nonumber \\
& + & \sum_{l=1}^{{n+n'-\vert\,n-n'\vert\over 2} -1}4\arctan\Bigl({x\over \vert\, n-n'\vert +2l}\Bigl)\Bigr\} \, ,
\label{Theta}
\end{eqnarray}
$n, n' = 1,...,(L-2S)/2$, and the number of pairs reads $(L-2S)/2=\infty$ in the 
TL provided that $(1-m_S)$ is finite. Moreover,
\begin{equation}
q_j = {2\pi\over L}I_j^n \, , \hspace{0.25cm} j=1,...,M_n^b \, ,
\label{qj}
\end{equation}
are the momentum values of a {\it $n$-band} associated with the set of $M_n$ $n$-pair configurations with the
same $n$ value and the quantum numbers $I_j^n$ 
are successive integers or half-odd integers according to the boundary conditions,
\begin{eqnarray}
I_j^n & = & 0,\pm 1,...,\pm {M_n^b -1\over 2} \, , \hspace{0.5cm} M_n^b\hspace{0.2cm}{\rm odd} \, ,
\nonumber \\
& = & \pm 1/2,\pm 3/2,...,\pm {M_n^b -1\over 2} \, , \hspace{0.5cm} M_n^b\hspace{0.2cm}{\rm even} \, .
\label{Ijn}
\end{eqnarray}
(Often an index $\alpha = 1,...,M_n$ is used to label the sub-set of occupied numbers $I_{\alpha}^n$ \cite{Taka}.)

For each $n$, there is a BA branch momentum $n$-band whose momentum values $q_j$, Eq. (\ref{qj}), are
such that $q_{j+1}-q_{j}={2\pi\over L}$ and have only occupancies zero and one. A $n$-band has $M_n^b=M_n+M^h_{n}$ 
such momentum values, $M_n$ of which are occupied by a single $n$-pair configuration. 
We call the occupied momentum values $n$-band particles. The $M^h_{n}$ 
momentum values left over are unoccupied. Here \cite{Taka},
\begin{equation}
M^h_{n} = 2S+\sum_{n'=n+1}^{\infty}2(n'-n)M_{n'} \, .
\label{Mh}
\end{equation}
We call such unoccupied momentum values $n$-band holes. Below we shall use the variables, 
\begin{eqnarray}
m_S & = & 2S/L\geq m \, , \hspace{0.50cm} m_n = M_n/L \, , 
\nonumber \\
m_n^h & = & M_n^h/L \, , \hspace{0.50cm} m_n^b = M_n^b/L \, .
\label{mmmm}
\end{eqnarray} 

The $n$-band momentum distribution function $M_n (q_j)$ (or $M_n^h (q_j)\equiv 1 - M_n (q_j)$) 
appearing in Eq. (\ref{gen-Lambda}) is such that $M_n (q_j)=1$ and $M_n (q_j)=0$ (or $M_n^h (q_j)=0$ and $M_n^h (q_j)=1$) for 
occupied and unoccupied values, respectively. Each LWS has specific values for that distribution.
The corresponding $n$-band discrete momentum variable, Eq. (\ref{qj}), has the range
$q_j \in [-q_n^b,q_n^b]$ where $q_n^b = \pi\,\left(m_n^b-{1\over L}\right)$, which in the TL simplifies to $q_n^b = \pi\,m_n^b$.
 
The momentum operator eigenvalues read,
\begin{equation}
P = \sum_{n=1}^{\infty}\sum_{j=1}^{M_n^b}M_{n} (q_{j})\,k_n (q_j) \, .
\label{P-k}
\end{equation}
where $k_n (q_j)$ is given in Eq. (\ref{kn-gen-Lambda}). By the use in this expression of
Eq. (\ref{gen-Lambda}), one finds that for all energy and momentum eigenstates
the summations $\sum_{n=1}^{\infty}\sum_{j=1}^{M_n^b}$ over the second term on the
right-hand side of that equation vanish, so that the momentum expression, Eq. (\ref{P-k}), 
simplifies to,
\begin{equation}
P = \sum_{n=1}^{\infty}\sum_{j=1}^{M_n^b}M_{n} (q_{j})\,q_j \, .
\label{P}
\end{equation}
That the momentum eigenvalues have this simple form
confirms that the quantum numbers $q_j$, Eq. (\ref{qj}), play the role of $n$-band
momentum values.

Other physical quantities such as the energy and the spin current of an energy and momentum eigenstate \cite{Taka}
also depend on the $n$-band momentum occupancy configurations through the
$n$-band momentum distribution functions $M_{n} (q_{j})$. However, in contrast to the
simple momentum expression, Eq. (\ref{P}), the dependence of the energy and spin current  
spectra on the occupied $\alpha = 1,...,M_n$ $n$-band momentum values $q_{j}=q_{\alpha}$ 
occurs through that of the rapidities real part $\Lambda_j^n = \Lambda^n (q_j)$ 
in Eq. (\ref{gen-Lambda}) and thus of $k_n (q_j)=2\arctan (\Lambda^n (q_j))$, Eq. (\ref{kn-gen-Lambda}).

Our study involves specifically the spin current $\langle\hat{J} (S)\rangle\equiv \langle l_{\rm r},S,-S\vert\hat{J} \vert l_{\rm r},S,-S\rangle$.
For any spin-$S$ LWS it is of the general form, 
\begin{equation}
\langle\hat{J} (S)\rangle=\sum_{n=1}^{\infty}\sum_{j =1}^{M_n^b} M_n (q_j)\,j_n (q_j) \, .
\label{J-part}
\end{equation}
Here the elementary currents are given by,
\begin{equation}
j_n (q_j) = -2J\,f_n (k_n (q_j)) \, .
\label{jn-fn}
\end{equation}
They involve the function,
\begin{equation}
f_n (k) = {2\over n}{\cos^2 (k/2)\over 2\pi\rho_n^b (\Lambda)}\sin k \, ,
\label{fn}
\end{equation}
whose variable $k\in [-\pi,\pi]$ is the parameter $k_n (q_j) = 2\arctan (\Lambda_j^n/n)$, 
Eq. (\ref{kn-gen-Lambda}). The variable of the function,
\begin{equation}
2\pi\rho_n^b (\Lambda)=2\pi\rho_n (\Lambda)+2\pi\rho_n^h (\Lambda) \, ,
\label{rhonb}
\end{equation}
in Eq. (\ref{fn}) is thus $\Lambda = n\tan (k/2)$ and $2\pi\rho_n (\Lambda)$ and $2\pi\rho_n^h (\Lambda)$ are 
the usual BA distributions \cite{Taka}. 

Solving the BA equations, Eq. (\ref{gen-Lambda}), gives the LWS set of $j=1,...,M_n^b$
real parameters $\Lambda_j^n = \Lambda^n (q_j)$  and  $k_n (q_j) = 2\arctan (\Lambda^n (q_j))$ both for occupied and 
unoccupied $n$-band momentum values. The former determine the actual energy and momentum eigenstate $j_n (q_j)$ values, Eq. (\ref{jn-fn}), 
in the spin current, Eq. (\ref{J-part}). 

The spin current expression, Eq. (\ref{J-part}), refers to a $n$-band particle current representation. 
As justified below, its alternative $n$-band hole representation reads,
\begin{equation}
\langle\hat{J}(S)\rangle = \sum_{n=1}^{\infty}{2S\over M_{n}^h} \sum_{j =1}^{M_n^b} M_n^h (q_j)\,j_n^h (q_j) \, ,
\label{J-f}
\end{equation}
where, 
\begin{equation}
j_n^h (q_j) \equiv - j_n (q_j) = 2J\,f_n (k_n (q_j)) \, ,
\label{jnh}
\end{equation}
and $j_n (q_j)$ is the elementary current, Eq. (\ref{jn-fn}). Such a $n$-band hole representation is physically
advantageous, as it has a more direct relation to the degrees of freedom of the 
above two sets of $L-2S$ and $2S$ original spins $1/2$, respectively.  

An $n$-band is exotic in that its momentum width, $2\pi\,m_n^b$, depends on
the value of $2S$ and the occupancies of the set of the $n'$-bands such that $n'\geq n$. Indeed,
$m_n^b=M_n^b/L$ where $M_n^b=M_n+M^h_{n}$ and $M^h_{n} = 2S+\sum_{n'=n+1}^{\infty}2(n'-n)M_{n'}$, Eq. (\ref{Mh}).
Hence, in contrast to the usual bands, the elementary current sum $\sum_{j =1}^{M_n^b} j_n (q_j)$ over all $n$-band momentum values
does not in general vanish. Actually, as found in Sec. \ref{D-algebra},
$\langle\hat{J} (0)\rangle \equiv \langle l_{\rm r} ,0,0\vert\hat{J}\vert l_{\rm r},0,0\rangle = 0$ 
and thus $\sum_{n=1}^{\infty}\sum_{j =1}^{M_n^b} M_n (q_j)\,j_n (q_j) =0$ for all $S=0$ LWSs 
implies that $\sum_{j =1}^{M_n^b} j_n (q_j)$ vanishes, provided that $\sum_{n'=n+1}^{\infty}2(n'-n)M_{n'}=0$
and thus $M^h_{n} = 2S$. Indeed, the usual relation $\sum_{j =1}^{M_n^b} j_n (q_j)=0$ is 
replaced by,
\begin{eqnarray}
\sum_{j =1}^{M_n^b} j_n (q_j) & = & \sum_{j =1}^{M_n^b}\sum_{n'=n+1}^{\infty}{2(n'-n)M_{n'}\over M_n^h}M_n^h (q_j)\,j_n (q_j) 
\nonumber \\
& = & {M_n^h-2S\over M_n^h}\sum_{j =1}^{M_n^b} M_n^h (q_j)\,j_n (q_j) \, ,
\label{sum-all}
\end{eqnarray}
which justifies the current general form, Eq. (\ref{J-f}).

To dig deeper into the physical meaning of the $n$-band hole representation,
we emphasize that the current, Eq. (\ref{J-f}), can be rewritten as,
\begin{equation}
\langle\hat{J} (S)\rangle = \sum_{n=1}^{\infty}\sum_{j =1}^{M_n^b} M_n^h (q_j)\,j_n^h (q_j)
+ \sum_{n=2}^{\infty}M_n\, j^p_n \, ,
\label{J-BA}
\end{equation}
where,
\begin{equation}
j^p_n = - \sum_{n'=1}^{n-1}{2 (n-n')\over M_{n'}^h}
\sum_{j=1}^{M^b_{n'}} M_{n'}^h (q_j)\,j_{n'}^h (q_j) \, .
\label{jpn}
\end{equation}

Now the degrees of freedom of the $2S$ spins $1/2$ that contribute to the currents
and those of the $L-2S$ that do not are distributed over the $\sum_{n=1}^{\infty}M_{n}^h$ $n$-band holes 
with elementary currents $j_n^h (q_j)$, Eq. (\ref{jnh}), the first term on the right-hand side of Eq. (\ref{J-BA}). 

The second term contains degrees of freedom of the $\sum_{n=2}^{\infty}n\,M_n=(L-2S)/2 - M_1$ spin-singlet pairs that
are bound within $n$-pair configurations with $n>1$ pairs, each carrying 
an elementary current $j^p_n$, Eq. (\ref{jpn}), whereas $M_1$ such pairs remain unbound.
For $S=0$, LWSs the latter currents exactly cancel 
those of the $n$-band holes. For $S>0$, LWSs they cancel the part of the $n$-band hole currents that
is not associated with the $2S$ spin-$1/2$ spins that contribute to the state's current. 

\section{Optimization of the spin current}
\label{optimizing}

The energy eigenstates that span each reduced subspace with fixed values for the spin $S$ 
and set of $n$-pair configuration numbers $\{M_n\}$ obeying the subspace spin-singlet pair
sum rule $\sum_{n=1}^{\infty}n\,M_n =(L-2S)/2$, Eq. (\ref{Nsingletpairs}),
are generated by configurations within which a number $M_n$ of momentum 
values $q_j = q_{\alpha} = {2\pi\over L}I_{\alpha}^n$ where $\alpha=1,...,M_n$ are occupied and 
the remaining $M^h_{n}$ momentum values are unoccupied. For each $n$-band
this gives a dimension ${M_n^b\choose M_n}$. 

Importantly, the value of the number $M_n^h = M_n^b - M_n$ of $n$-band holes that naturally emerges from the TBA, Eq. (\ref{Mh}),
ensures that for each $S$-fixed subspace the dimension, Eq. (\ref{NsingletS}), can alternatively be written as \cite{Taka},
\begin{equation}
{\cal{N}}_{singlet} (S) = \sum_{\{M_{n}\}}\,\prod_{n =1}^{\infty} {M_n^b\choose M_n} \, .
\label{Nsinglet-MM}
\end{equation}
Here $\sum_{\{M_{n}\}}$ is a summation over all sets of $\{M_{n}\}$ obeying the exact fixed-$S$
number of spin-singlet pairs sum rule $\sum_{n=1}^{\infty}n\,M_n=(L-2S)/2$, Eq. (\ref{Nsingletpairs}).
Each spin-$S$ subspace can thus be divided into a set of the above smaller reduced subspaces with 
fixed values for the set of $n$-pair configuration numbers $\{M_n\}$ obeying the sum rule. 

That, as confirmed by Eq. (\ref{Nsinglet-MM}), all ${\cal{N}}_{singlet} (S)$ independent spin-singlet configurations (Eq. (\ref{NsingletS})) 
of each $S$-fixed subspace are contained within the sets of $\{M_{n}\}$ $n$-pair configurations obeying the fixed-$S$ sum 
rule $\sum_{n=1}^{\infty}n\,M_n=(L-2S)/2$, Eq. (\ref{Nsingletpairs}), is consistent with the $n$ Bethe-ansatz rapidities, Eq. (\ref{Lambda-jnl-ideal}),
being indeed associated with $n$ spin-singlet pairs. 

We denote the maximum spin current expectation value $\langle l_{\rm r} ,S,-S\vert\hat{J}\vert  l_{\rm r},S,-S\rangle$
in each such subspaces by ${\cal{J}}_{\rm max} (m_S,\{m_n\})$. 
Here $n>1$, since $m_1={1\over 2}(1-m_S-\sum_{n=2}^{\infty}2n\,m_n)$ is uniquely determined. 
That maximum value refers to a LWS whose $n$-bands have
occupancies that maximize their contribution to the spin current,
Eqs. (\ref{J-f}) and (\ref{J-BA}). 

Let $M_n^{\rm max} (q_j)$ where $n=1,...,(L-2S)/2$
denote the $n$-band momentum distribution functions of the energy eigenstate 
that carries the current ${\cal{J}}_{\rm max} (m_S,\{m_n\})$. Then there is
in the same subspace another energy eigenstate whose $n$-band momentum distribution 
functions are given by $M_n^{\rm min} (q_j)=M_n^{\rm max} (-q_j)$ (and
$M_n^{\rm min} (0)=M_n^{\rm max} (0)$ if the quantum numbers $I_j^n$, Eqs. (\ref{qj}) and (\ref{Ijn}),
are integers.) It carries a current ${\cal{J}}_{\rm min} (m_S,\{m_n\})=-{\cal{J}}_{\rm max} (m_S,\{m_n\})$
whose absolute value is also maximum, $\vert{\cal{J}}_{\rm min} (m_S,\{m_n\})\vert=\vert{\cal{J}}_{\rm max} (m_S,\{m_n\})\vert$.
Hence the same maximum current absolute value is reached by a current optimization 
that refers to a current maximization or minimization, respectively. 

If our goal were the very involved problem of calculating the stiffness for all temperatures $T>0$,
a detailed numerical analysis of the Bethe-ansatz solutions would be required. However, the
primary aim of this paper is calculating an exact stiffness upper bound to clarify whether in the TL
the stiffness vanishes or is finite as $m\rightarrow 0$. This makes the problem studied in the following
technically simpler than the involved calculations needed to
access the stiffness for all temperatures $T>0$. 

We start by identifying which of the
current absolute values $\vert{\cal{J}}_{\rm max} (m_S,\{m_n\})\vert$ of all reduced subspaces with the same
$S$ value and different sets of $\{M_{n}\}$ obeying the sum rule $\sum_{n=1}^{\infty}n\,M_n=(L-2S)/2$, 
Eq. (\ref{Nsingletpairs}), is the largest. Concerning the corresponding current optimization, in the following
we account for the current maximization and minimization both reaching exactly the same maximum current absolute value.
Fortunately, this task shows basic similarities to that of finding the ground-state energy and thus minimizing 
the Bethe-ansatz energy in each subspace spanned by energy eigenstates with fixed $S^z$ 
and $\{M_n\}$ values \cite{Taka}.

In both cases one finds that the smallest energy (and largest spin current absolute value) refers
to subspaces spanned by energy and momentum eigenstates that are not populated by $n$-pair configurations
with $n>1$ bound spin-singlet pairs. These two types of states are populated by $M_1= (L-2S)/2$ unbound
spin-singlet pairs, the corresponding $n=1$ momentum band having compact $q_j$ occupancies.
(The ground-state compact occupancies are for $S=0$ and $S=L/2$ given in the two unnumbered
equations, respectively, appearing just below Eq. (2.12) of Ref. \cite{Taka}.)
The only difference is that the resulting ground-state compact $n=1$ momentum band occupancy is symmetrical,
$M_{1} (q_j) - M_{1} (-q_j)=0$, or quasi-symmetrical, $M_{1} (q_j) - M_{1} (-q_j)=\pm2\pi/L$,
whereas as given in Eqs. (\ref{msmaller3}) and (\ref{mlargerer3}) of Appendix \ref{opti} that of the maximum spin 
current absolute value is fully asymmetrical. 

As a first step of the above program, one straightforwardly confirms from the use of the TBA equations (Eq. (\ref{gen-Lambda}))
in Eqs. (\ref{J-f}) - (\ref{jpn}) that for each of the reduced 
subspaces spanned by energy eigenstates with fixed $S$ and $\{M_n\}$ values
the $n$-band $q_j$ occupancies that maximize the spin current are indeed asymmetric 
and compact. Specifically, if (a) $m_n \geq m_n^h$ or (b) $m_n^h \geq m_n$ the $n$-band
occupancy that maximizes its contribution to $|\langle l_{\rm r} ,S,-S\vert\hat{J}\vert  l_{\rm r},S,-S\rangle|$
refers to an asymmetric compact distribution of $q_j>0$ or $q_j<0$ momentum values,
respectively, (a) with width $2\pi\,m_n^h$ of all $M_n^h$ holes or (b) with width $2\pi\,m_n$ of all $M_n$ 
$n$-band particles. For these occupancy configurations, 
Eqs. (\ref{gen-Lambda}) simplify to
\begin{eqnarray}
& & k_n (q_j) = q_j 
\nonumber \\
& + & \left(\sum_{q_{j'}=-q_n^b}^{q_{j_0^h}}+\sum_{q_{j'}=q_{j_0^h}+2\pi m_n^h}^{q_n^b}\right)
\Theta_{n\,n'} (\Lambda_j^n-\Lambda_{j'}^{n'}) \, ,
\label{Lambda-compcA}
\end{eqnarray}
for $q_{j'}\in [q_{j_0^h},q_{j_0^h}+2\pi m_n^h]\geq 0$ and $M_n \geq M_n^h$ and to,
\begin{equation}
k_n (q_j) = q_j + \sum_{q_{j'}=q_{j_0}}^{q_{j_0}+2\pi m_n}
\Theta_{n\,n'} (\Lambda_j^n-\Lambda_{j'}^{n'}) \, ,
\label{Lambda-compcB}
\end{equation}
for $q_{j'}\in [q_{j_0},q_{j_0}+2\pi m_n]\leq 0$ and $M_n \leq M_n^h$. In these equations,
$ k_n (q_j)$ and $\Theta_{n\,n'} (x)$ are the functions, Eqs. (\ref{kn-gen-Lambda}) and (\ref{Theta}), respectively,
$n , n' = 1,...,(L-2S)/2$, $q_{j_0^h}=q_n^b-2\pi m_n^h$, and $q_{j_0}\in [-q_n^b,-\pi/2+\pi m_n]$ changes from
$q_{j_0}=-q_n^b$ for finite $m_n=m_n^h$ to $q_{j_0}=-\pi/2+\pi m_n$ for small $m_n<m_n^h$.

The exact stiffness upper bound expression derived in Sec. \ref{two} is analytical, the
same applying to the corresponding largest maximum spin current absolute value used in its calculation.
Hence for simplicity and as in the identification of the ground state and corresponding subspace minimum energy 
\cite{Taka}, in the following and in Appendices \ref{string} and \ref{opti} we skip intermediate technicalities and focus on the description of
the the main steps of our calculations and corresponding physical meaning in terms of spin-singlet pairs configurations.

As mentioned above, the current with largest maximum absolute value
is for all $S>0$ values carried by an energy eigenstate 
whose singlet configurations refer to $(L-2S)/2$ unbound spin $1/2$ pairs. 
Importantly, energy eigenstates for which all $(L-2S)/2$ spin-singlet pairs are unbound are described only by
real rapidities, Eq. (\ref{Lambda-jnl-ideal}) for $n=1$. The resulting largest current maximum absolute value is 
separated by a current gap from the maximum absolute values of currents carried by energy and momentum eigenstates 
populated by bound spin-singlet pairs. 

In the following we introduce the smallest and largest 
maximum spin current absolute values $\vert{\cal{J}}_{\rm max} (m_S,\{m_n\})\vert$ and the two corresponding reduced subspaces
for each fixed $S$ value. In Appendix \ref{opti} we address the issue 
of the intermediate maximum spin current absolute values and corresponding current gap of the remaining reduced subspaces with the same $S$ value.

From the use of Eqs. (\ref{Lambda-compcA}) and (\ref{Lambda-compcB}) in the current expressions, 
Eqs. (\ref{J-f}) - (\ref{jpn}), one finds that the smallest $\vert{\cal{J}}_{\rm max} (m_S,\{m_n\})\vert$ value is reached for
the reduced subspace for which $M_n=1$ for $n =(L-2S)/2\geq 2$ and $M_n=0$ for $n<(L-2S)/2$. For it
the solution of Eqs. (\ref{Lambda-compcA}) and (\ref{Lambda-compcB}) leads to,
\begin{eqnarray}
\Lambda_j^n & = & n\tan (q_j/2) \, ,
\nonumber \\
2\pi\rho_n^b (\Lambda) & = & 2\pi\rho_n (\Lambda)+2\pi\rho_n^h (\Lambda)  
= {2\over n}{1\over 1+ (\Lambda/n)^2} \, ,
\nonumber \\
q_j & = & 0,\pm {2\pi\over L},...,\pm {2\pi\over L}\,(S -1),\pm {2\pi\over L}\,S \, .
\label{L-minJ}
\end{eqnarray}
For this reduced subspace, the current maximum absolute value is achieved when the single $n$-band particle
occupies the momentum $q_j=-q_n^b=-{2\pi\over L}\,S$ for $S\leq L/4$ and $q_j=-\pi/2$
for $L/4\leq S\leq L/2-2$. This gives,
\begin{eqnarray}
\vert{\cal{J}}_{\rm max} (m_S,m_n)\vert & = & 2J\sin (\pi m_S) \, , \hspace{0.5cm} S\leq L/4 \, ,
\nonumber \\
& = & 2J \, , \hspace{0.5cm} L/4\leq S\leq L/2-2 \, ,
\label{minJ}
\end{eqnarray}
for $n=(L-2S)/2$. 

In general $\vert{\cal{J}}_{\rm max} (m_S,m_n)\vert $ is an ${\cal{O}} (L)$ object,
in contrast to its smallest value, Eq. (\ref{minJ}). Indeed the current maximum absolute value
of each reduced subspace decreases when the $(L-2S)/2$ spin-singlet pairs 
become less diluted relative to the remaining $2S$ original spins $1/2$.
The less diluted case refers specifically to all $(L-2S)/2$ spin-singlet pairs being bound within a single $n$-pair configuration.
This refers to present subspace, for which the current maximum absolute value, Eq. (\ref{minJ}), is smallest.

On the other hand, the largest current maximum absolute value $\vert{\cal{J}}_{\rm max} (m_S,\{m_n\})\vert$ is reached 
in the opposite limit. It refers to the reduced subspace for which $M_1 = (L-2S)/2$ and $M_n=0$ for $n>1$ and thus $M_1^h = 2S$. For it
all $(L-2S)/2$ spin-singlet pairs remain unbound. The energy and momentum eigenstate carrying this current is a superposition 
of local configurations within which the $(L-2S)/2$ spin-singlet pairs are most diluted relative to the remaining $2S$ 
original spins $1/2$.

Since $q_{j+1}-q_{j}={2\pi\over L}$ and we are interested in the $L\gg 1$ limit, in the following analysis the set of successive 
momentum values $\{q_{j}\}$ is replaced by a corresponding continuum momentum variable, $q\in [-q_1^b,q_1^b]$. Hence 
the set of real rapidities $\Lambda^1_j = \Lambda^1 (q_j)$ becomes a function $\Lambda^1 (q)$ of the continuum momentum 
variable $q$. That for this reduced subspace $M_n=0$ for $n>1$, simplifies the problem to a single equation, which is
Eq. (\ref{Lambda-compcA}) or Eq. (\ref{Lambda-compcB}) for $n=1$. 

One of the two energy and momentum eigenstates in such a reduced subspace
that carry spin currents with the same maximum absolute value is for each $S$ value associated with the compact and asymmetrical
$n=1$ band distribution function given in Eqs. (\ref{msmaller3}) and (\ref{mlargerer3}) of Appendix \ref{opti}.
The use of such a distribution function in Eq. (\ref{Lambda-compcA}) and Eq. (\ref{Lambda-compcB}) for $n=1$
leads within the continuum momentum variable representation to,
\begin{eqnarray}
& & 2\arctan (\Lambda^1 (q)) = q + {1\over 2\pi}\left(\int_{-q_1^b}^{q_{j_0^h}}+\int_{q_{j_0^h} + 2\pi m_S}^{q_1^b}\right) dq'
\nonumber \\
& \times & 2\arctan \left({\Lambda^1 (q)-\Lambda^1 (q')\over 2}\right)  \, , 
\label{L1qA}
\end{eqnarray} 
for $m_S \leq 1/3$ and,
\begin{eqnarray}
& & 2\arctan (\Lambda^1 (q)) = q + {1\over 2\pi}\int_{q_{j_0}}^{q_{j_0} + \pi (1 -m_S)}dq' 
\nonumber \\
& \times & 2\arctan \left({\Lambda^1 (q)-\Lambda^1 (q')\over 2}\right) \, , 
\label{L1qB}
\end{eqnarray} 
for $m_S \geq 1/3$, respectively. Here the value of $\Theta_{n\,n'} (x)$, Eq. (\ref{Theta}), for $n=n'=1$ was used. 

Solution of Eq. (\ref{L1qA}) or (\ref{L1qB}) uniquely defines the function $\Lambda^1 (q)$. Moreover, the distributions 
$2\pi\rho_1 (\Lambda)$ and $2\pi\rho_1^h (\Lambda)$ and thus
$2\pi\rho_1^b (\Lambda) = 2\pi\rho_1 (\Lambda) + 2\pi\rho_1^h (\Lambda)$ are defined by the equation,
\begin{eqnarray}
2\pi\rho_1^b (\Lambda) & = & {2\over 1+\Lambda^2}
\nonumber \\
& - & {1\over 2\pi}\left(\int_{-\infty}^{B_0^h} + \int_{B_S^h}^{\infty}\right)d\Lambda'\,{2\pi\rho_1 (\Lambda')\over 1+\left({\Lambda -\Lambda'\over 2}\right)^2} \, ,
\label{2pisugmaA}
\end{eqnarray} 
for $m_S \leq 1/3$ and,
\begin{equation}
2\pi\rho_1^b (\Lambda) = {2\over 1+\Lambda^2}
- {1\over 2\pi}\int_{B_0}^{B_S}d\Lambda'\,{2\pi\rho_1 (\Lambda')\over 1+\left({\Lambda -\Lambda'\over 2}\right)^2} \, , 
\label{2pisugmaB}
\end{equation} 
for $m_S \geq 1/3$. In these equations,
\begin{eqnarray}
B_0^h & = & \Lambda^1 (q_{j_0^h}) 
\, ; \hspace{0.5cm} B_S^h = \Lambda^1 (q_{j_0^h} + 2\pi m_S) \, ,
\nonumber \\
B_0 & = & \Lambda^1 (q_{j_0})
\, ; \hspace{0.5cm} B_S= \Lambda^1 (q_{j_0} + \pi (1 -m_S)) \, .
\nonumber 
\end{eqnarray} 

The largest current maximum absolute value uniquely defined by the solutions of Eqs. (\ref{L1qA}) or (\ref{L1qB}) and 
Eqs. (\ref{2pisugmaA}) or (\ref{2pisugmaB}), respectively, is of the form,
\begin{equation}
\vert{\cal{J}}_{\rm max} (m_S)\vert\equiv \vert{\cal J}_{\rm max}(m_S,\{m_n=0\})\vert = J C(m_S) L \, ,
\label{J-mn-0}
\end{equation}
where the coefficient $C(m_S)$ is a $L$-independent function of $m_S\in [0,1]$ with a single maximum 
at an intermediate $m_S$ value. It vanishes both in the $m_S\rightarrow 0$ and $m_S\rightarrow 1$ limits,
the corresponding limiting behaviors being,
\begin{eqnarray}
C(m_S) & = & {\pi\over 4}m_S \, , \hspace{0.5cm} m_S \ll 1 \, ,
\nonumber \\
& = & (1-m_S) \, , \hspace{0.5cm} (1-m_S) \ll 1 \, .
\label{CmS}
\end{eqnarray} 
Except in these limits, for the present asymmetric BA quantum numbers distributions 
the current maximum absolute value at fixed $S$, Eq. (\ref{J-mn-0}), is an ${\cal O}(L)$ object.

On the other hand, the quantity needed for the derivation of the stiffness upper bound 
is rather the corresponding ratio $\vert{\cal{J}}_{\rm max} (m_S)\vert/2S$, which is independent of $L$
and smoothly decreases as $m_S$ increases. Specifically, it decreases from its $m_S\ll 1$ maximum value, $J\pi/4$, 
to $J(1-m_S)$ for $(1-m_S)\ll 1$, vanishing as $m_S\rightarrow 1$. 

Since $\vert{\cal{J}}_{\rm max} (m_S)\vert/2S$ is a decreasing function of $m_S$, it turns out that
the value of the parameter $C(m_S)$ on the right-hand side of Eq. (\ref{J-mn-0}) that is relevant for our goal of 
clarifying whether the stiffness upper bound derived in the following vanishes or remains finite as $m\rightarrow 0$ 
is $C(m_S) = {\pi\over 4}m_S$, which is reached for $m_S \ll 1$, Eq. (\ref{CmS}).

As discussed in Appendix \ref{opti}, all reduced subspaces other than the two considered in this section are associated with 
$\vert{\cal{J}}_{\rm max} (m_S,\{m_n\})\vert$ values that obey the inequalities,
\begin{eqnarray}
\vert{\cal{J}}_{\rm max} (m_S,\{m_n\}) & \geq & \vert{\cal{J}}_{\rm max} (m_S,m_n)\vert_{n=(L-2S)/2}
\nonumber \\
\vert{\cal{J}}_{\rm max} (m_S,\{m_n\}) & \leq & \vert\leq\vert{\cal{J}}_{\rm max} (m_S)\vert \, .
\label{in-gen-J}
\end{eqnarray}
Hence an important quantity for our studies is the current gap $\Delta_J =\Delta_J (m_S,\{m_n\})$ defined as,
\begin{equation}
\Delta_J = \vert{\cal{J}}_{\rm max} (m_S)\vert-\vert{\cal{J}}_{\rm max} (m_S,\{m_n\})\vert\geq 0 \, .
\label{gap}
\end{equation}
It separates the current maximum absolute values $\vert{\cal{J}}_{\rm max} (m_S)\vert$ and $\vert{\cal{J}}_{\rm max} (m_S,\{m_n\})\vert$, respectively, 
whose currents are carried by energy and momentum eigenstates 
with the same $S$ value and different sets of $\{M_{n}\}$ obeying the sum rule $\sum_{n=1}^{\infty}n\,M_n=(L-2S)/2$, 
Eq. (\ref{Nsingletpairs}). (We recall that in the case of $\vert{\cal{J}}_{\rm max} (m_S)\vert$ this refers to a state with $M_1=(L-2S)/2$ and $M_n=0$
for $n>1$.) 

In Appendix \ref{opti} we discuss the mechanism that determines for the whole range $m_S \in [0,1]$ the occurrence 
of the current gap, Eq. (\ref{gap}), between the largest current maximum absolute value and the maximum current 
absolute values of all other reduced subspaces with the same spin $S$. For $(1-m_S)\ll 1$ the following analytical
expression is found in Appendix \ref{opti} for such a current gap,
\begin{equation}
\Delta_J = 2J\sum_{n=1}^{\infty} (n-1)\,M_n \, , \hspace{0.5cm} (1-m_S)\ll 1 \, .
\label{gap-mS-1}
\end{equation}

In general the current gap is finite, $\Delta_J >0$. As justified in Appendix \ref{opti}, the only exception 
occurs in the $m_S\rightarrow 0$ limit and thus also as $m\rightarrow 0$, in which $\Delta_J \rightarrow 0$ for 
currents ${\cal{J}}_{\rm max} (m_S,\{m_n\})$ of energy eigenstates 
for which $\sum_{n=2}^{\infty}n\,M_n >0$ but $\sum_{n=2}^{\infty}n\,m_n \rightarrow 0$ as $L\rightarrow\infty$. 
As discussed in that Appendix, the minimum value of the current gap is an increasing function of $m_S$ with
limiting behaviors,
\begin{eqnarray}
{\rm min}\,\Delta_J & = & 0 \, , \hspace{0.5cm} m_S \rightarrow 0 \, ,
\nonumber \\
& = & 2J \, , \hspace{0.5cm} m_S \rightarrow 1 \, .
\label{lim-min-gap}
\end{eqnarray}

Another important current gap is,
\begin{equation}
\Delta_J^0 = \vert{\cal{J}}_{\rm max} (m_S)\vert-\vert{\cal{J}} (m_S)\vert\geq 0 \, .
\label{gap0}
\end{equation}
It separates the absolute values of currents ${\cal{J}}_{\rm max} (m_S)$ and ${\cal{J}} (m_S)$ 
of energy and momentum eigenstates with the same fixed $S$ value and without $n$-pair configurations with $n>1$ bound pairs. 
Now there are energy eigenstates other than that carrying the maximum current ${\cal{J}}_{\rm max} (m_S)$ 
for which $\Delta_J^0$ is an ${\cal{O}} (1/L)$ object such that $\Delta_J^0\rightarrow 0$ as $L\rightarrow\infty$.
Moreover, as discussed above there is one energy eigenstate that carries the minimum current
${\cal{J}}_{\rm min} (m_S)=-{\cal{J}}_{\rm max} (m_S)$ for which
$\Delta_J^0 = \vert{\cal{J}}_{\rm max} (m_S)\vert-\vert{\cal{J}}_{\rm min} (m_S)\vert = 0$. 

Our derivation of an exact stiffness upper bound relies on both the inequalities $\Delta_J\geq 0$ and
$\Delta_J^0\geq 0$ holding for all $S$ values. We believe that there is some symmetry protecting the current 
gap $\Delta_J$, so that the inequality $\Delta_J\geq 0$ found in this paper is always obeyed. 
The validity of that inequality is, in the $L\rightarrow\infty$ limit, independent of the corresponding $n$-pair configurations 
fine structure, Eq. (\ref{Lambda-jnl-ideal}) and Eq. (\ref{Lambda-jnl}) of Appendix \ref{string}.
Indeed, the stiffness upper bound derived in this paper is insensitive to the values of the spin currents carried 
by energy and momentum eigenstates populated by $n>1$ bound spin-singlet pairs. 
Those involve Bethe ansatz complex rapidities, Eq. (\ref{Lambda-jnl-ideal}), 
whose imaginary parts describe the binding of the $n>1$ spin-singlet pairs within the $n$-pair configurations.  

That in the $m\rightarrow 0$ limit the minimum value of the gap $\Delta_J$ behaves as $\Delta_J\rightarrow 0$
shows that only within it the maximum absolute values $\vert{\cal{J}}_{\rm max} (m_S,\{m_n\})\vert$ 
of the spin currents carried by some of the energy and momentum eigenstates populated by $n>1$ bound spin-singlet pairs barely equals
the largest maximum absolute value $\vert{\cal{J}}_{\rm max} (m_S)\vert = J C(m_S) L = J C(m) L $, Eq. (\ref{J-mn-0}), which reads
$J L {\pi\over 4}m$ in that limit. This does not affect though the stiffness upper bound as computed in Sec. \ref{two}.

The effects of the finite-system string deviations \cite{Vladimirov-84,Essler-92,Isler-93,Takahashi-03,Caux-07}
on the spin currents and corresponding current gap $\Delta_J$ is a problem discussed in Appendix \ref{string}. Such deviations
only affect the complex rapidities, Eq. (\ref{Lambda-jnl-ideal}) and Eq. (\ref{Lambda-jnl}) of Appendix \ref{string}
for $n>1$, associated with the binding of the spin-singlet pairs within a $n$-pair configuration. 
Indeed, such $n$-pair configurations prevail, only the fine structure of their rapidity strings deviate from the ideal form,
Eq. (\ref{Lambda-jnl-ideal}). 

Furthermore, in the $L\rightarrow\infty$ limit the energy eigenstates that carry the maximum spin currents
remain those whose spin-singlet pairs are unbound, which are not affected by the finite-system string deviations.
This is consistent with in the $L\rightarrow\infty$ limit such deviations not contributing to the 
thermodynamics provided that $T>0$ and $m\neq 0$ \cite{Caux-07,Wiegmann-83}.

\section{Two exact spin stiffness upper bounds}
\label{two}

A general result of our above analysis is that energy eigenstates that are a superposition of local configurations within which the 
$(L-2S)/2$ spin-singlet pairs are more diluted relative to the remaining $2S$ original spins $1/2$ carry larger currents. 
The maximum dilution is achieved by energy eigenstates with $M_1=(L-2S)/2$ unbound
pairs and no bound pairs, which for the finite system are described by undeformed real rapidities. That for $L\rightarrow\infty$ such states absolute maximum 
current value, $\vert{\cal{J}}_{\rm max} (m_S)\vert=J C(m_S) L$, Eq. (\ref{J-mn-0}), is exact follows from its independence 
for $L\rightarrow\infty$ from the complex rapidity strings fine structure, Eq. (\ref{Lambda-jnl-ideal}) and Eq. (\ref{Lambda-jnl}) of Appendix \ref{string}. 

Consistently, the reduced-subspace 
current maximum absolute value has its smallest value, Eq. (\ref{minJ}), in the opposite limit in which all $(L-2S)/2$ spin-singlet pairs are bound
within a single $n$-pair configuration, {\it i.e.} $M_n=1$ with $n=(L-2S)/2$. 

In the intermediate general situation corresponding to reduced subspaces for which the $(L-2S)/2$ spin-singlet pairs are distributed by $n$-pair
configurations with two or more different $n$ values such that $\sum_{n=1}^{\infty}n\,M_n = (L-2S)/2$,
the corresponding maximum current absolute value always obeys the inequalities, Eq. (\ref{in-gen-J}).

Except in the $m_S\rightarrow 0$ and $m_S\rightarrow 1$ limits, for the present asymmetric BA quantum numbers distributions 
the largest current maximum absolute value at fixed $S$, 
$\vert{\cal{J}}_{\rm max} (m_S)\vert=J C(m_S) L$, Eq. (\ref{J-mn-0}), is an ${\cal O}(L)$ object.  
A first exact stiffness upper bound, 
\begin{equation}
D^*_{S^z} (T) = {(2S^z)^2\over 2 L T}\sum_{S=\vert S^z\vert}^{L/2}\sum_{l_{\rm r}} 
p_{l_{\rm r},S,S^z}\left({{\cal{J}}_{\rm max} (m_S)\over 2S}\right)^2 \, .
\label{D-all-T-simp-12}
\end{equation}
is obtained by replacing the moduli of the expectation values $\vert\langle l_{\rm r},S,-S\vert\hat{J} \vert l_{\rm r},S,-S\rangle\vert$ 
of LWSs with the same $S$ value in the stiffness expression, Eq. (\ref{D-all-T-simp}),
by their largest maximum values $\vert{\cal{J}}_{\rm max} (m_S)\vert$ of each $S$-fixed subspace. 

The ratio $\vert{\cal{J}}_{\rm max} (m_S)\vert/2S$ appearing in Eq. (\ref{D-all-T-simp-12}) is 
independent of $L$. It smoothly decreases upon increasing $m_S$ from its $m_S\ll 1$ maximum value, $J\pi/4$, 
to $J(1-m_S)$ for $(1-m_S)\ll 1$. For each fixed-$S^z$ canonical ensemble,
the largest ratio $\vert{\cal{J}}_{\rm max} (m_S)\vert/2S$ in the $S$ summation of Eq. (\ref{D-all-T-simp-12})  
is then that referring to the minimum $S$ value, $S=\vert S^z\vert = m\,L/2$, such that $m_S = m$.
 
A second exact stiffness upper bound is obtained by replacing in Eq. (\ref{D-all-T-simp-12}) the ratio $\vert{\cal{J}}_{\rm max} (m_S)\vert/2S$
by its largest value $\vert{\cal{J}}_{\rm max} (m)\vert/2\vert S^z\vert=\vert{\cal{J}}_{\rm max} (m)\vert/(m\,L)$. 
Importantly, the state summations in Eq. (\ref{D-all-T-simp-12}) can then be performed exactly for all finite temperatures $T>0$. Indeed, the probability 
distribution $p_{l_{\rm r},S,S^z}$ in each fixed-$S^z$ canonical ensemble is normalized as 
$\sum_{S=\vert S^z\vert}^{L/2}\sum_{l_{\rm r}} p_{l_{\rm r},S,S^z}=1$. Such state summations account for the 
subspace dimensions and thus as well for the full Hilbert-space dimension, $\sum_{2S=0\,({\rm integers})}^{L}\,{\cal{N}}(S) = 2^{L}$.
For $T>0$ the resulting (larger) upper bound $D^{**}_{S^z} \geq D^*_{S^z} \geq D_{S^z}$, then becomes, 
\begin{equation}
D^{**}_{S^z} (T) = {{\cal{J}}_{\rm max}^2 (m)\over 2 T L} \, .
\label{D**}
\end{equation}

From the use of Eqs. (\ref{J-mn-0}) and (\ref{CmS}) one finds that for $m\ll 1$ its value is,
\begin{equation}
D^{**}_{S^z} (T) = {\left(J{\pi\over 4}\right)^2\,m^2\,L\over 2T} \, , \hspace{0.25cm} m \ll 1 \, ,
\label{value-D-m0-1}
\end{equation}
so that $D\leq \left(J{\pi\over 4}\right)^2\,m^2\,L/(2T)$ for $T>0$ and $m\ll 1$ determining the constant $c$ in the 
upper bound, Eq. (\ref{Dcm0}), to be $c=(J \pi/4)^2/2$. 

Moreover, one finds again from the use of Eqs. (\ref{J-mn-0}) and (\ref{CmS}), 
\begin{equation}
D^{**}_{S^z} (T) = {J^2\,(1-m)^2\,L\over 2T} \, , \hspace{0.25cm} (1-m)\ll 1 \, ,
\label{value-D-m1}
\end{equation}
so that $c'=J^2/2$ in Eq. (\ref{Dclinem1}). 

This completes our proof of the vanishing spin stiffness in the 
TL, $L\to\infty$, for any fixed range or even distribution of $S^z$, or any distribution of $m$ shrinking sufficiently 
fast that $\langle m^2 \rangle L \to 0$. 

Note that ${\cal{J}}_{\rm max}^2 (m)/L \leq (J\pi/4)^2\,m^2\,L$ for all $m\in [0,1]$. Hence we may use the small-$m$ 
upper-bound expression in Eq. (\ref{value-D-m0-1}) for the 
whole $m\in [0,1]$ range, which is the bound stated in the abstract. 	

\section{Concluding remarks}
\label{concluding}

At $\Delta =0$ the exact value of the high-temperature spin stiffness of the spin-$1/2$ $XXZ$ chain whose Hamiltonian
is given in Eq. (\ref{HXXZ}) is $D = (J/4)^2/T$, so that the lower bound of Ref. \cite{Tomaz-11} saturates it. Our exact result that 
at $\Delta=1$ and both $m=0$ and $m\rightarrow 0$ the spin stiffness vanishes within a canonical ensemble at all finite temperatures applies to
high temperature as well. This implies that the above lower bound saturates the high-temperature spin stiffness both at $\Delta =0$
and $\Delta =1$. Combined with the equality of that lower bound to the TBA spin stiffness found in Ref. \cite{Zotos-99} at 
$\lambda =\pi/l'$, this most likely implies that the bound saturates the stiffness for the whole range $0\leq \Delta\leq 1$.
This provides also strong evidence that the divergences emerging in the integrands of Eqs. (24) and (25) of Ref. \cite{ANI-05} 
at $m=0$ cancel each other in the case of systems whose stiffness is finite at $T=0$, as in the case of the XXZ chain at $m=0$ for 
$0\leq \Delta\leq 1$.

That the spin stiffness of the spin-$1/2$ $XXZ$ chain vanishes at the isotropic point in the TL
is an exact result that refers to the canonical ensemble. It leaves out, marginally, the grand canonical ensemble
in which $\langle m^2\rangle = {\cal O}(1/L)$. However, the large overestimate of the elementary BA currents we used in deriving 
the stiffness upper bound, Eq. (\ref{D**}), leads us to expect that our result remains valid in the grand canonical case, in accord
with the usual expectation of the equivalence of ensembles in the TL.

%%%%%%%%%%%%%%%%%%%%%%%%%%%%%%%%%%%%%%%%%%%%%%%%%%%%%%%%%%%%%%%%%%%%%%%%%%
\acknowledgements
We thank Henrik Johannesson, Pedro D. Sacramento, Steven R. White and Xenophon Zotos for discussions. J. M. P. C. thanks the support by the Beijing CSRC and the 
Portuguese FCT through the Grant PEST-C/FIS/UI0607/2013. T. P. acknowledges support by Slovenian ARRS Grant No. P1-0044. D. K. C. acknowledges the hospitality 
of the International Institute of Physics at the Universidade Federal do Rio Grande do Norte in Natal, Brazil, where part of this work 
was conducted.

%%%%%%%%%%%%%%%%%%%%%%%%%%%%%%%%%%%%%%%%%%%%%%%%%%%%%%%%%%%%%%%%%%%%%%%%%%

\appendix

\section{Effects of the finite-system bound pairs string deformations}
\label{string}

Here we discuss the effects of the fine structure of the finite-system $n>1$ bound pairs 
complex-rapidity string deformations. 
We consider the general Bethe-ansatz equations, Eq. (\ref{gen-Lambda-BA}), for large but finite chains.
The corresponding general spin current expectation values read,
\begin{equation}
\langle\hat{J} (S)\rangle = \sum_{\alpha} \,j (q_{\alpha}) \, ,
\label{J-S}
\end{equation}
where again the summation is over occupied $q_{\alpha}$ values and the elementary currents $j (q_j)$ are given by,
\begin{eqnarray}
j (q_j) & = & -2J\, {2(\cos (k_j/2))^2\over 2\pi\rho^b (\Lambda_j)} \sin k_j \, ,
\nonumber \\
k_j & = & 2\arctan (\Lambda_j) \, .
\label{j-k}
\end{eqnarray}
Here $2\pi\rho^b (\Lambda_j)=2\pi\rho (\Lambda_j)+2\pi\rho^h (\Lambda_j)$,
$2\pi\rho (\Lambda_j)$, and $2\pi\rho^h (\Lambda_j)$ are the usual Bethe-ansatz  distributions \cite{Taka}.

Only some of the solutions of Eq. (\ref{gen-Lambda-BA}) are in terms of real rapidities.
There also exist solutions involving groups of complex rapidities. Some of the latter arrange themselves into deformed 
strings \cite{Vladimirov-84,Essler-92,Isler-93,Takahashi-03,Caux-07}, which 
deviate from the ideal complex-rapidity strings, Eq. (\ref{Lambda-jnl-ideal}).
Specifically, the roots of Eq. (\ref{gen-Lambda-BA}) are partitioned in a configuration of strings, where a $n$-string is a group
of $n$ roots such that,
\begin{equation}
\Lambda_j^{n,l} = \Lambda_j^n + i (n+1-2l) + D_j^{n,l} \hspace{0.5cm} l = 1,...,n \, .
\label{Lambda-jnl}
\end{equation}
From comparison with Eq. (\ref{Lambda-jnl-ideal}) one finds that the only difference refers
to the deviation $D_j^{n,l} = R_j^{n,l} + i \delta_j^{n,l}$ where $R_j^{n,l}$ and $\delta_j^{n,l}$ 
are real numbers. (Self-conjugacy implies that $D_j^{n,l} = (D_j^{n,n+1-l})^*$.) 
For large but finite chains such deformed strings change 
the spin currents, Eqs. (\ref{J-S}) and (\ref{j-k}), carried by the corresponding complex-rapidity ideal-string energy eigenstates. 

For the real-rapidity energy eigenstates considered in Section \ref{optimizing},
Eqs. (\ref{L1qA}) and (\ref{L1qB}) remain the same if one uses the general Bethe-ansatz equation, Eq. (\ref{gen-Lambda-BA}). Indeed, one finds from the
use of Eq. (\ref{Lambda-jnl}) with $n=l=1$ in the general Bethe ansatz equation, Eq. (\ref{gen-Lambda-BA}),
that $D_j^{1,1} = 0$. Hence the real rapidities are not deformed for large but finite chains.

For $n=1$ the rapidity, Eq. (\ref{Lambda-jnl}), is real and refers to a single unbound pair. The imaginary part that emerges 
for $n>1$ in Eq. (\ref{Lambda-jnl}) is still associated with the binding of the corresponding $n$ spin-singlet pairs.
Indeed, the $n$-pair configurations prevail, the only change being the deviation $D_j^{n,l}$ in their fine structure. 
The TBA equations, Eq. (\ref{gen-Lambda}), are recovered by using $D_j^{n,l}=0$ in Eq. (\ref{Lambda-jnl}). 

Various authors have studied the fine structure of the deformed string solutions
for finite chains \cite{Vladimirov-84,Essler-92,Isler-93,Takahashi-03,Caux-07}, which are associated with the deviations $D_j^{n,l}$ in Eq. (\ref{Lambda-jnl}). 
In the case of two-string solutions, it is found that there are narrow and
wide branches \cite{Vladimirov-84,Essler-92,Isler-93,Takahashi-03}. The wide deformed strings lie on a curve in the complex plane. Their imaginary
part diverges upon increasing the real part. On the other hand, the narrow strings become closer
to the real line with increasing real part. They finally collapse onto it, so that no narrow string
solutions occur for high quantum numbers. Instead, extra solutions appear with two real roots.
The investigations of Ref. \cite{Caux-07} address higher string cases. They show that the
collapse of narrow pairs both from two-strings and from higher strings is the only aberration
from the ideal strings, Eq. (\ref{Lambda-jnl-ideal}), if one allows for deviations $D_j^{n,l}$ in the strings 
themselves, Eq. (\ref{Lambda-jnl}).

It is widely accepted that as far as thermodynamics are concerned the use of complex-rapidity
ideal strings gives for $L\rightarrow\infty$ correct results as long as the temperature and spin density 
are not strictly vanishing \cite{Caux-07,Wiegmann-83}. This applies to the stiffness upper bound derived in this paper, 
Eq. (\ref{D**}), which refers to $T>0$ and $m>1$. Deformed peripheral strings whose center increases with $L$ do not contribute to 
thermodynamics \cite{Caux-07,Wiegmann-83}. The irrelevance in the $L\rightarrow\infty$ limit of the finite-system 
deformations of the strings that contribute to thermodynamics can be checked directly \cite{Caux-07}. Previous 
studies have considered deformed strings for large but finite chains of form, Eq. (\ref{Lambda-jnl}),
keeping $\Lambda_j^n$ approximately fixed. They have found that these deformations decrease exponentially 
with increasing $L$ \cite{Takahashi-03,Caux-07}. (See figure 11 of Ref. \cite{Caux-07}.) 

Our analysis of the effects of the finite-system deformed strings simplifies because it focuses only
on whether in the $L\rightarrow\infty$ limit the current maximum absolute value of deformed strings 
energy and momentum eigenstates  
overcome the current gap $\Delta_J$, Eq. (\ref{gap}), that separates the maximum spin current absolute 
value $\vert{\cal{J}}_{\rm max} (m_S)\vert$, Eq. (\ref{J-mn-0}), from the absolute values of the 
spin currents carried by complex-rapidity states within the ideal strings, Eq. (\ref{Lambda-jnl-ideal}). 

We have confirmed that the use of the $L\rightarrow\infty$ behavior of the deformed-string solutions found by other authors 
\cite{Vladimirov-84,Essler-92,Isler-93,Takahashi-03,Caux-07} in the spin current expression, Eqs. (\ref{J-S}) and (\ref{j-k}),
reaches the same results as using in such an expression the finite large-$L$ deformed-string solutions and 
taking the $L\rightarrow\infty$ limit in the end. Both such procedures reveal that the largest currents remain those
carried by energy eigenstates that are a superposition of local configurations within which the $(L-2S)/2$ spin-singlet pairs are most diluted 
relative to the remaining $2S$ original spins $1/2$. 

The finite-system string deformations affect only the bound spin-singlet pairs. Consistently, the finite-system deformations do not prevent 
the energy eigenstates whose configurations involve $M_1=(L-2S)/2$ unbound pairs from carrying 
the spin current with largest absolute value, Eq. (\ref{J-mn-0}). Indeed, the maximum dilution is achieved when all $(L-2S)/2$ 
spin-singlet pairs remain unbound. Such states have no bound spin-singlet pairs and are described by 
real rapidities only. As mentioned above, in large but finite chains those remain undeformed. 

Specifically, from the use of Eq. (\ref{gen-Lambda-BA}) in the current expressions, Eqs. (\ref{J-S}) and (\ref{j-k}),
one finds that  energy and momentum eigenstates such that $\sum_{n=1}^{\infty}2n\,M_n/L$ remains finite and $\sum_{n=1}^{\infty}M_n/L$ vanishes as 
$L\rightarrow\infty$ carry currents that remain finite as $L\rightarrow\infty$. Hence their absolute values are
much smaller than $\vert{\cal{J}}_{\rm max} (m_S)\vert$, Eq. (\ref{J-mn-0}), which is an ${\cal{O}}(L)$ object.
Furthermore, we find that the largest currents carried by complex-rapidity  energy and momentum eigenstates with deformed strings 
refer to string lengths $n$ and occupancies $M_n$ such that $\sum_{n=2}^{\infty}2n\,M_n/L$ vanishes in the $L\rightarrow\infty$ limit.
The solution of the general Bethe-ansatz equation, Eq. (\ref{gen-Lambda-BA}), for such states is obtained by 
expanding it around that of Eqs. (\ref{L1qA}) and (\ref{L1qB}) for the real-rapidity state that carries current with largest absolute value. The minimum 
value of the current gap $\Delta_J$ is then found to remain finite for $S>0$. This result is not affected by the finite-system collapse of narrow pairs. 
Moreover, in the $L\rightarrow\infty$ limit the minimum current gap still changes from zero for 
$m_S\rightarrow 0$ to $2J$ for $(1-m_S)\ll 1$, Eq. (\ref{lim-min-gap}). 

Interestingly, in the $L\rightarrow\infty$ limit the general current gap expression accounting for string
deformations remains  $2J\sum_{n=1}^{\infty} (n-1)\,M_n$, Eq. (\ref{gap-mS-1}), when $(1-m_S)\ll 1$ and thus $\sum_{n=1}^{\infty}2n\,M_n/L\ll 1$.
This result is again consistent with larger currents  being carried by energy eigenstates that are a 
superposition of local configurations within which the $(L-2S)/2$ spin-singlet pairs are more diluted 
relative to the remaining $2S$ original spins $1/2$.

We then conclude that independent of the fine structure of the fnite-system 
complex-rapidity energy eigenstates strings, at fixed $S$ and in the $L\rightarrow\infty$ 
limit the current of largest maximum absolute value is carried by the energy eigenstate considered in Section \ref{optimizing}
whose $(L-2S)/2$ spin-singlet pairs are all unbound. For a finite but large chain such a state is described by undeformed real rapidities.

\section{Reduced subspaces current maximum absolute values and corresponding current gap $\Delta_J$}
\label{opti}

In this Appendix we address the issue of the intermediate maximum spin current absolute values $\vert{\cal{J}}_{\rm max} (m_S,\{m_n\})\vert$ 
and corresponding current gap $\Delta_J$ of the set of reduced subspaces with the same $S$ value other than the two considered in 
Sec. \ref{optimizing}. Our discussion refers to the TBA associated with the
complex rapidities, Eq. (\ref{Lambda-jnl-ideal}). The effects in the TL of the finite-system deviations $D_j^{n,l}$ 
in Eq. (\ref{Lambda-jnl}) of Appendix \ref{string} from such complex rapidities is an issue discussed in
that Appendix.

From the use of Eqs. (\ref{Lambda-compcA}) and 
(\ref{Lambda-compcB}) in the current expressions, Eqs. (\ref{J-f}) - (\ref{jpn}),
one finds that for $0<m_S<1$ and reduced subspaces for which $\sum_{n=1}^{\infty}2n\,m_n$ remains 
finite and $\sum_{n=1}^{\infty}m_n$ vanishes as $L\rightarrow\infty$, the current gap, Eq. (\ref{gap}),
is an ${\cal{O}}(L)$ object. 

Furthermore, the current gap is finite and independent of $L$ for reduced subspaces 
for which both $\sum_{n=2}^{\infty}2n\,M_n$ and $\sum_{n=1}^{\infty}m_n$ are finite.
It is smallest for reduced subspaces for which $\sum_{n=2}^{\infty}2n\,M_n$
is finite and $\sum_{n=2}^{\infty}2n\,m_n$ vanishes in the $L\rightarrow\infty$ limit.
For the latter reduced subspaces and $L\gg 1$, one has that the
inequality $\sum_{n=2}^{\infty}2n\,m_n\ll 2m_1$ holds for $(1-m_S)$ finite in,
\begin{eqnarray}
(1-m_S) = 2m_1 + \sum_{n=2}^{\infty}2n\,m_n \, .
\nonumber
\end{eqnarray}
Moreover, the inequality $\sum_{n'=n+1}^{\infty}2(n'-n)\,m_{n'}\ll m_S$ holds for $m_S$ finite and $L\gg 1$ in, 
\begin{eqnarray}
m_n^h = m_S + \sum_{n'=n+1}^{\infty}2(n'-n)\,m_{n'} \, .
\nonumber
\end{eqnarray}

One can then expand the solution of Eqs. (\ref{Lambda-compcA}) and (\ref{Lambda-compcB}) for these subspaces around that
of Eqs. (\ref{L1qA}) and (\ref{L1qB}). This involves $n$-band momentum distribution functions of the form
$M_{n} (q_j) = M_{n}^0 (q_j) + \delta M_{n} (q_j)$ where $M_{n}^0 (q_j) =0$ for $n>1$ and $M_{1}^0 (q_j)$
is the $n=1$ band momentum distribution function used in Eqs. (\ref{L1qA}) and (\ref{L1qB}). It reads,
\begin{eqnarray}
M_{1}^0 (q_j) & = & 1 \hspace{0.25cm}{\rm for}\hspace{0.2cm}q_j \in [-(2\pi m_S-q_1^b), q_1^b] 
\nonumber \\
& = & 0 \hspace{0.25cm}{\rm for}\hspace{0.2cm}q_j \in [-q_1^b,-(2\pi m_S-q_1^b)] \, ,
\label{msmaller3}
\end{eqnarray}
for $m_S\leq 1/3$ where $(2\pi m_S-q_1^b)\geq 0$ and,
\begin{eqnarray}
M_{1}^0 (q_j) & = & 1 \hspace{0.25cm}{\rm for}\hspace{0.2cm}q_j \in [-q_{j0}, -(q_{j0} + \pi (1-m_S))]
\nonumber \\
& = & 0 \hspace{0.25cm}{\rm for}\hspace{0.2cm}q_j \in [-q_1^b,-q_{j0}] 
\nonumber \\
& = & 0 \hspace{0.25cm}{\rm for}\hspace{0.2cm}q_j \in [-(q_{j0} - \pi (1-m_S)),q_1^b] \, ,
\label{mlargerer3}
\end{eqnarray}
for $m_S\geq 1/3$ where $(q_{j0} + \pi (1-m_S))\geq 0$. 

The $n=1$ band momentum distribution function, Eqs. (\ref{msmaller3}) and (\ref{mlargerer3}),
is for each $S$ value associated with an energy and momentum eigenstate that carries a spin current whose absolute value, Eq. (\ref{J-mn-0}), 
is largest. For that state the $n=1$ band limiting momentum $q_1^b$ appearing in Eqs. (\ref{msmaller3}) and (\ref{mlargerer3}) reads,
\begin{equation}
q_1^b={\pi\over 2}\left(1-m_S-{2\over L}\right)\approx {\pi\over 2}\left(1-m_S\right) \, ,
\label{q1bJmax}
\end{equation}
and the momentum $q_{j0}> 0$ is in Eq. (\ref{mlargerer3}) a continuous decreasing function of $m_S \in [1/3,1]$ that changes from 
$q_{j0}={\pi\over 2}\left(1+m_S\right)={2\pi\over 3}$ for $m_S=1/3$ to $q_{j0}={\pi\over 2}m_S={\pi\over 2}$ at $m_S=1$.

As discussed in the following, the current deviations associated with the $n$-bands momentum distribution function deviations
$\delta M_{n} (q_j)$ in $M_{n} (q_j) = M_{n}^0 (q_j) + \delta M_{n} (q_j)$ are such that the minimum value of the current gap, Eq. (\ref{gap}), 
increases from zero for $m_S\rightarrow 0$ to $2J$ for $m_S\rightarrow 1$, as given in Eq. (\ref{lim-min-gap}).

The larger current maximum absolute values of reduced subspaces spanned by energy eigenstates involving
complex rapidities and thus populated by bound spin-singlet pairs, $\vert{\cal{J}}_{\rm max} (m_S,\{m_n\})\vert \leq \vert{\cal{J}}_{\rm max} (m_S)\vert$,
refer for the whole $m_S \in [0,1]$ range to subspaces 
for which $\sum_{n=2}^{\infty}2n\,M_n$ is finite but $\sum_{n=2}^{\infty}2n\,m_n$ vanishes as $L\rightarrow\infty$.
For such reduced subspaces and $m_S\ll 1$, the elementary currents $j_n^h (q_j)$, Eq. (\ref{jnh}),
on the right-hand side of Eqs. (\ref{J-f}) and (\ref{J-BA}) vanish for $n>1$ as $L\rightarrow\infty$ whereas the
$n=1$ elementary currents $j_1^h (q_j)$ are finite. 

We now justify why the minimum value of the current gap, Eq. (\ref{gap}), vanishes as $m_S\rightarrow 0$.
From the use in the current expression, Eq. (\ref{J-f}), of the
compact asymmetric distributions that maximize the currents of the 
$m_S\ll 1$ reduced subspaces we then find,
\begin{equation}
\vert{\cal{J}}_{\rm max} (m_S)\vert  = {L\over 2\pi}\int_{{\pi\over 2}(1+m_S)-2\pi m_S}^{{\pi\over 2}(1+m_S)}dq\,{\rm j}_1^h (q) \approx J{\pi\over 4} 2S \, ,
\label{J-f-0}
\end{equation}
for the $\sum_{n=2}^{\infty}2n\,M_n=0$ subspace and,
\begin{eqnarray}
\vert{\cal{J}}_{\rm max} (m_S,\{m_n\})\vert  & = & {2S\over M_{1}^h} \times {L\over 2\pi}\int_{\pi m_1^b -2\pi m_1^h}^{\pi m_1^b}dq\,{\rm j}_1^h (q) 
\nonumber \\
& \approx & 
{2S\over M_{1}^h} \times J{\pi\over 4}  M_{1}^h = J{\pi\over 4} 2S \, ,
\label{J-f-max-mSmn}
\end{eqnarray}
for subspaces for which $\sum_{n=2}^{\infty}2n\,M_n$ is finite and $\sum_{n=2}^{\infty}2n\,m_n$ vanishes as $L\rightarrow\infty$.
The $n=1$ elementary currents appearing in these equations read ${\rm j}_1^h (q) = j_1^h (q) + \delta j_1^h (q) = J(\pi/4)\sin q$ where
$\delta j_1^h (q)$ stands for a deviation from the bare elementary currents $j_1^h (q)$, Eq. (\ref{jnh}) for $n=1$, that stems from phase
shifts. Moreover, the numbers $m_1^b$ and $m_1^h$ in Eq. (\ref{J-f-max-mSmn}) read
$m_1^b = {1\over 2}(1+m_S+\sum_{n=3}^{\infty}2(n-2)\,m_n)$ and  $m_1^h=m_S+\sum_{n=2}^{\infty}2(n-1)\,m_n$, respectively.

Since for the present $m_S\ll 1$ reduced subspaces one has that $j_n^h (q)=0$ for $n>1$,
the main effect on the current maximum absolute values of reduced subspaces spanned by energy eigenstates involving
complex rapidities is in this limit that the number of $n=1$ band holes increases 
from $M_1^h=2S$ to $M_1^h=2S+\sum_{n=2}^{\infty}2(n-1)\,M_n$. 

On the one hand (and as follows 
from analysis of Eqs. (\ref{J-f-0}) and (\ref{J-f-max-mSmn}), except for the factor $2S/M_{1}^h$ in the latter equation),
for $S$ finite and thus $m_S\rightarrow 0$ as $L\rightarrow\infty$ this effect increases the contribution
to the current maximum absolute value from $J(\pi/4)\,2S$ to $J(\pi/4)\,M_1^h = J(\pi/4)\,(2S+\sum_{n=2}^{\infty}2(n-1)\,M_n)$.

On the other hand, the current cancelling term on the right-hand side of Eq. (\ref{J-BA}) leads to the
factor $2S/M_{1}^h$ in the current expressions, Eqs. (\ref{J-f}) and (\ref{J-f-max-mSmn}). 
In the present limit, this factor exactly cancels the increase in the current of the $\sum_{n=2}^{\infty}2n\,M_n>0$ subspaces
due to the enhancement in the number of $n=1$ band holes, as confirmed from comparison of Eqs.  (\ref{J-f}) and (\ref{J-f-0}).
This is why the minimum current gap vanishes for $m_S\rightarrow 0$.

Upon increasing $m_S$ within its domain $m_S\in [0,1]$, the elementary currents $j_n^h (q_j)$, Eq. (\ref{jnh}), 
become finite for $n>1$ but their maximum absolute value remains in general smaller than that of the $n=1$ elementary currents $j_1^h (q_j)$.
The current canceling term on the right-hand side of Eq. (\ref{J-BA}) vanishes for the $\sum_{n=2}^{\infty}2n\,M_n=0$
energy and momentum eigenstate with current maximum absolute value whereas it is finite for the 
$\sum_{n=2}^{\infty}2n\,M_n>0$ energy and momentum eigenstates with current maximum absolute value. For
$m_S\rightarrow 0$ it is behind the factor $2S/M_{1}^h$ in Eq. (\ref{J-f-max-mSmn}), which renders the currents of Eqs. (\ref{J-f-0}) 
and (\ref{J-f-max-mSmn}) equal. However, upon increasing $m_S$ it leads only to a partial cancelation, the net result being that 
the maximum absolute value of the overall spin current carried by the $\sum_{n=2}^{\infty}2n\,M_n>0$ energy and momentum eigenstates 
becomes smaller than that carried by the $\sum_{n=2}^{\infty}2n\,M_n=0$ energy and momentum eigenstate with the same $S$ value. 

That the minimum value of the current gap $\Delta_J$, Eq. (\ref{gap}), increases upon increasing $m_S$ is simple to realize 
for $m_S > 1/3$. Then the number of $n=1$ band particles becomes smaller
than that of $n=1$ band holes. For the range $1/3<m_S<1$ it is then more convenient to use the representation within which the $n$-band 
particles are the effective current carriers. Within that representation the spin currents are given by Eq. (\ref{J-part}).
For $1/3<m_S<1$ the maximum absolute values of the corresponding $n>1$ elementary currents $j_n (q_j)$, Eq. (\ref{jn-fn}),
appearing in Eq. (\ref{J-part}) remain in general smaller than that of the $n=1$ elementary currents $j_1 (q_j)$. 

Furthermore, the use of the number
of pairs sum rule $\sum_{n=1}^{\infty}n\,M_n = (L-2S)/2$, Eq. (\ref{Nsingletpairs}), reveals that one has $M_1 = M_1^0 = (L-2S)/2$ 
effective current carriers for the $\sum_{n=2}^{\infty}2n\,M_n=0$ reduced subspace 
whereas that number decreases to $\sum_{n=1}^{\infty}M_n<(L-2S)/2=\sum_{n=1}^{\infty}n\,M_n$ for
the other reduced subspaces. Hence $M_1^0 = (L-2S)/2$ for the former reduced subspace exactly equals 
$\sum_{n=1}^{\infty}n\,M_n = (L-2S)/2$ for the latter subspaces. The difference in the number of effective current carriers of both types
of reduced subspaces then reads,
\begin{eqnarray}
\delta N_{\rm carriers} & = & M_1^0 - \sum_{n=1}^{\infty}M_n = \sum_{n=1}^{\infty}n\,M_n - \sum_{n=1}^{\infty}M_n
\nonumber \\
& = & \sum_{n=1}^{\infty} (n-1)\,M_n \, .
\label{M10}
\end{eqnarray}

On the one hand, each of the $M_1 = M_1^0 = (L-2S)/2$ $n=1$ band particles of the reduced subspace for which 
$M_n =0$ for $n>1$ carry an elementary current $j_1 (q_j)$, Eq. (\ref{jn-fn}) for $n=1$, whose absolute value is in general larger than 
that of the $n>1$ elementary currents $j_n (q_j)$ in the general spin current expression, Eq. (\ref{J-part}). The latter are
carried by the $n>1$ band particles of subspaces with a finite number $\sum_{n=2}^{\infty}M_n$ of $n$-pair configurations with $n>1$ pairs. 

On the other hand, the number $M_1 = M_1^0 = (L-2S)/2$ of current carriers of
the subspace for which $M_n =0$ for $n>1$ is larger than the number $\sum_{n=1}^{\infty}M_n<(L-2S)/2$ of carriers of  
subspaces with a finite number of $n>1$ $n$-pair configurations. Indeed, $(L-2S)/2=\sum_{n=1}^{\infty}n\,M_n>\sum_{n=1}^{\infty}M_n$
for the latter subspaces. 

The interplay of these two properties justifies why the reduced subspace with no $n>1$ band particles
is that whose maximum current absolute value is largest.
This effect is easiest to describe in the $(1-m_S)\ll 1$ limit in which the minimum value of the current gap $\Delta_J$, Eq. (\ref{gap}), 
reaches its largest value, $\Delta_J = 2J$. Within this limit the $n>1$ elementary currents $j_n (q_j)$, Eq. (\ref{jn-fn}), in 
Eq. (\ref{J-part}) become equal to the $n=1$ elementary currents $j_1 (q_j)$. Specifically, $j_n (q_j) = -2J\sin q_j$ 
for all $n$ values. (In this limit there are no phase-shift elementary current deviations, $\delta j_n (q_j) =0$.) 

The asymmetric compact distribution of the $M_n$ $n$-band particles 
is for $(1-m_S)\ll 1$ centred at momentum $q_j=-\pi/2$. Combining that for $(1-m_S)\ll 1$ and thus $m_n\ll 1$ all $n$-band particles 
carry the same elementary current $\approx -2J\sin (-\pi/2) = 2J$ with the number 
$\sum_{n=1}^{\infty}M_n$ of $n$-band particles at fixed $S$ value decreasing 
for larger occupancies of $n$-bands with increasingly larger number $n$ of bound
spin-singlet pairs, one finds that in that limit the general current gap $\Delta_J$, Eq. (\ref{gap}), reads $2J\sum_{n=1}^{\infty} (n-1)\,M_n$. 

This is indeed the exact expression, Eq. (\ref{gap-mS-1}), obtained in this limit from the use of 
the solutions of Eqs. (\ref{Lambda-compcA}) and (\ref{Lambda-compcB}) in 
the current expressions, Eqs. (\ref{J-f}) - (\ref{jpn}).
It equals $2J$ times the difference in the number of effective current
carriers $\delta N_{\rm carriers}$ of both types of reduced subspaces, Eq. (\ref{M10}).  
The minimum value of the $m_S\rightarrow 1$ current gap $\Delta_J$, Eq. (\ref{gap-mS-1}), is $2J$. It corresponds to the reduced subspace 
for which $M_1 = (L-2S)/2 - 2$, $M_2 =1$, and $M_n=0$ for $n>2$. 

Consistently, the minimum value of the general current gap $\Delta_J$, Eq. (\ref{gap}), 
is in the range $m_S \in [0,1]$ an increasing a function of $m_S$ with limiting values given in Eq. (\ref{lim-min-gap}).

\end{document}